\documentclass[10pt,journal,compsoc]{IEEEtran}

\ifCLASSOPTIONcompsoc
  \usepackage[nocompress]{cite}
\else
  \usepackage{cite}
\fi

\ifCLASSINFOpdf
  \usepackage[pdftex]{graphicx}
\else
  \usepackage[dvips]{graphicx}
\fi

\ifCLASSOPTIONcompsoc
  \usepackage[caption=false,font=footnotesize,labelfont=sf,textfont=sf]{subfig}
\else
  \usepackage[caption=false,font=footnotesize]{subfig}
\fi

\usepackage{fixltx2e}
\usepackage{stfloats}
\usepackage{float}

\usepackage{url}

\usepackage{bm}
\usepackage{amsfonts}
\usepackage{amssymb}
\usepackage{algorithmic}
\usepackage{amsmath}
\usepackage{graphicx}
\usepackage{enumitem}
\usepackage{array}
\usepackage{booktabs}	
\usepackage{widetable}
\usepackage{epstopdf}
\usepackage[normalem]{ulem}
\usepackage{multirow}
\usepackage{listings}
\usepackage{color}
\usepackage{caption}
\usepackage[T1]{fontenc}

\usepackage[ruled]{algorithm2e}

\usepackage{scalerel}
\usepackage{tikz}
\usetikzlibrary{svg.path}

\definecolor{orcidlogocol}{HTML}{A6CE39}
\tikzset{
  orcidlogo/.pic={
    \fill[orcidlogocol] svg{M256,128c0,70.7-57.3,128-128,128C57.3,256,0,198.7,0,128C0,57.3,57.3,0,128,0C198.7,0,256,57.3,256,128z};
    \fill[white] svg{M86.3,186.2H70.9V79.1h15.4v48.4V186.2z}
                 svg{M108.9,79.1h41.6c39.6,0,57,28.3,57,53.6c0,27.5-21.5,53.6-56.8,53.6h-41.8V79.1z M124.3,172.4h24.5c34.9,0,42.9-26.5,42.9-39.7c0-21.5-13.7-39.7-43.7-39.7h-23.7V172.4z}
                 svg{M88.7,56.8c0,5.5-4.5,10.1-10.1,10.1c-5.6,0-10.1-4.6-10.1-10.1c0-5.6,4.5-10.1,10.1-10.1C84.2,46.7,88.7,51.3,88.7,56.8z};
  }
}

\newcommand\orcidicon[1]{\href{https://orcid.org/#1}{\mbox{\scalerel*{
\begin{tikzpicture}[yscale=-1,transform shape]
\pic{orcidlogo};
\end{tikzpicture}
}{|}}}}

\usepackage{hyperref} 

\usepackage{color}
\definecolor{olive}{rgb}{0,0.4,0.2}
\definecolor{olivegreen}{rgb}{0.1,0.8,0.3}
\definecolor{mauve}{rgb}{0.48,0,0.72}

\newcommand{\AAAA}[1]{{\color{black}#1}\normalfont}
\newcommand{\AAA}[1]{{\color{black}#1}\normalfont}

\def\etal{\textit{et~al.}}

\begin{document}
%
\title{Rhythm is a Dancer: Music-Driven Motion Synthesis with Global Structure}
%
%
%
%

\author{Andreas~Aristidou$^{\textsuperscript{\orcidicon{0000-0001-7754-0791}}}$,~\IEEEmembership{Senior Member,~IEEE},
        Anastasios~Yiannakidis$^{\textsuperscript{\orcidicon{0000-0002-3721-6548}}}$,
        Kfir~Aberman$^{\textsuperscript{\orcidicon{0000-0002-4958-601X}}}$,
        Daniel~Cohen-Or$^{\textsuperscript{\orcidicon{0000-0001-6777-7445}}}$,
        Ariel~Shamir$^{\textsuperscript{\orcidicon{0000-0001-7082-7845}}}$,~\IEEEmembership{Member,~IEEE},
        and~Yiorgos~Chrysanthou$^{\textsuperscript{\orcidicon{0000-0001-5136-8890}}}$ 
\IEEEcompsocitemizethanks{\IEEEcompsocthanksitem A. Aristidou, A. Yiannakidis, and Y. Chrysanthou are with the Department of Computer Science, University of Cyprus, Nicosia, 1678, Cyprus, and the CYENS Centre of Excellence, Nicosia, 1016, Cyprus. \protect\\
E-mail: a.aristidou@ieee.org; tasyiann@gmail.com; yiorgos@cs.ucy.ac.cy}
\IEEEcompsocitemizethanks{\IEEEcompsocthanksitem K. Aberman and D. Cohen-Or are with the Department of Computer Science, Tel-Aviv University, Ramat Aviv, Tel-Aviv, 6997801, Israel. \protect\\
E-mail: kfiraberman@gmail.com, cohenor@gmail.com}
\IEEEcompsocitemizethanks{\IEEEcompsocthanksitem A. Shamir is  with the Department of Computer Science, The Interdisciplinary Center Herzliya (now Reichman University), 4610101, Israel. \protect\\
E-mail: arik@idc.ac.il}
\thanks{Manuscript received \today; revised \today.}}

%
%

\markboth{}%
{Aristidou \MakeLowercase{\textit{et al.}}: Rhythm is a Dancer}
%



\IEEEtitleabstractindextext{%
\begin{abstract}
Synthesizing human motion with a global structure, such as a choreography, is a challenging task. Existing methods tend to concentrate on local smooth pose transitions and neglect the global context or the theme of the motion.  In this work, we present a music-driven motion synthesis framework that generates long-term sequences of human motions which are synchronized with the input beats, and jointly form a global structure that respects a specific dance genre. In addition, our framework enables generation of diverse motions that are controlled by the content of the music, and not only by the beat.
Our music-driven dance synthesis framework is a hierarchical system that consists of three levels: \textit{pose}, \textit{motif}, and \textit{choreography}. The pose level consists of an LSTM component that generates temporally coherent sequences of poses. The motif level guides sets of consecutive poses to form a movement that belongs to a specific distribution using a novel \emph{motion perceptual-loss}. And the choreography level selects the order of the performed movements  and drives the system to follow the global structure of a dance genre. Our results demonstrate the effectiveness of our music-driven framework to generate natural and consistent movements on various dance types, having control over the content of the synthesized motions, and respecting the overall structure of the dance.
\end{abstract}

}

\maketitle

\IEEEdisplaynontitleabstractindextext

%
\IEEEpeerreviewmaketitle

\section{Introduction}
\label{section:Introduction}


\IEEEPARstart{D}{ance} is ``a performing-art form consisting of purposefully selected and controlled rhythmic sequences of human movements''~\cite{wikipedia} for an aesthetic purpose. Computationally synthesizing a dance is challenging not only because motions must be continuous, smooth and expressive locally, but also because a dance has a meaningful global temporal structure~\cite{Fraleigh:1987,Foster:2011}.
Recent advances in machine learning using neural networks have shown promising results in controlling articulated characters and creating arbitrary realistic human motions, including dance. However, current methods provide no means to control the structure of the synthesized motion globally. They tend to concentrate on local smooth motions and neglect the context or the global theme of the dance. 
These challenges become more acute when the goal is to synthesize and control the motion in response to a given music. A dancing pose at any moment can be followed by various possible poses, while the long-term spatio-temporal structure of body movements results in high kinematic complexity~\cite{Lee:2019:dancing2music}.

In this paper, we present a technique to generate dance motions that follow a given audio beat and music features, but also follows the contextual and cultural meaning of the dance, which we call \emph{choreography}.
Previous works in audio-driven human body animation built movement transition graphs that are synchronized to the beat~\cite{Kim:2003,Sauer:2009,Bellini:2018}, or the emotion~\cite{Shiratori:2006}, while more recent works use either hidden Markov models~\cite{Ofli:2012}, or recurrent neural networks~\cite{Alemi:2017,Shlizerman:2018,Tang:2018}. 
These methods generate motions that follow the given audio beat, while following a specific style, but show limited variability and lack  global consistency that is dictated by a meaningful dance context. In contrast, our technique generates long-term dance animations, within a natural and feasible set of movements, that are well aligned with a given musical style and rhythm, and adequate to the diversity of motions existing in a specific dance genre. In addition, the framework enables high-level choreography control of the dance content.


\begin{figure*}[t]
\centering
 \includegraphics[width=1\linewidth]{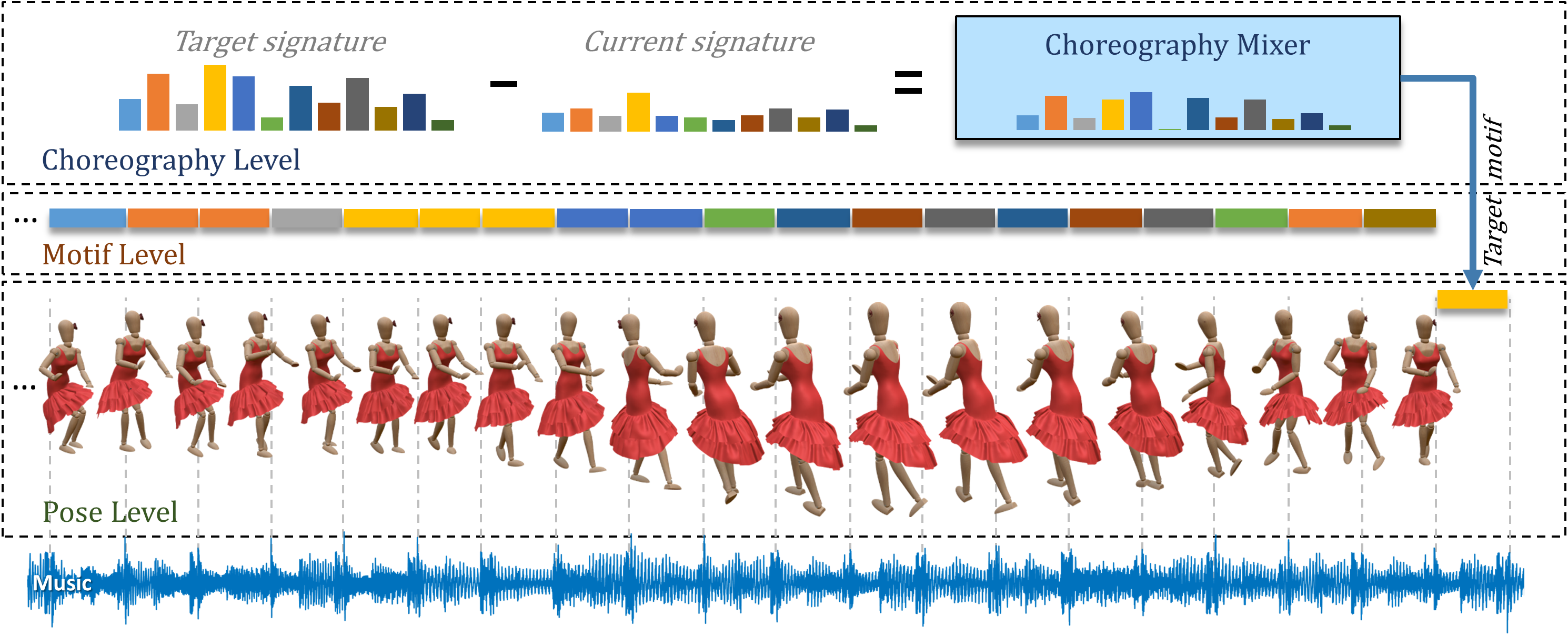}
 \caption{Music-driven motion synthesis. Our framework enables generation of rich and realistic human motions synchronized with music that form a global structure respecting the culture of a specific dance genre. Our framework is composed of three levels (bottom to top): (i) the pose level - that produces temporally coherent poses synchronized with beats, (ii) the motif level - that ensures the short movements belong to specific motion cluster (motifs) (iii) and the choreography level - that takes care of the global, long-term, structure of the dance, by selecting the next target motif to match the signature of the dance.}
\label{fig:teaser}
\end{figure*}

Our music-driven dance synthesis framework consists of three levels (see Figure~\ref{fig:teaser}): \emph{pose}, \emph{motif}, and \emph{choreography}. At the pose-level, we generate motion frames by employing an auto-conditional long short-term memory (acLSTM) network~\cite{Zhou:2018} that learns the temporal evolution of the motion in a specific dancing style. In contrast to the original acLSTM, our network is conditioned on audio features that enable the synthesized motion to follow the beat, and also on motion motifs that enable control over the global, high-level, structure of the motion.

We define a motion \emph{motif} similarly to~\cite{Aristidou:2018} as the centroid of a cluster of motion-words in some embedded space, where motion-words are short temporal sequences of consecutive poses that represent a movement. We use their pretrained neural network to convert the motion words to motion motifs.



The motif-level is responsible for the naturalness and style of small parts of the movement.
The poses between two consecutive beats are collected in this level and novel variations of the resulting motion words are guided by the audio features. These features drive an Adaptive Instance Normalization (AdaIN) layer that applies temporally invariant transformation to the synthesized motion words. A novel \emph{motion perceptual-loss} ensures that the resulting motion-word belongs to the driving motif.

The choreography-level controls the global structure of the dance. Dance is not merely a series of motion words. A dance has a structure, a theme, a culture, which we express by the distribution of motion motifs, or \emph{signature}. In each step, the network is guided to generate motion words so that the synthesized dance signature will match, in the long-term, the signature of the target dance. In addition, special care is given to temporal coherency by selecting motion motifs that match consecutively and create an appropriate sequence to drive the animation. 

We demonstrate the effectiveness of our music-driven framework on various dance genres, including salsa, modern, and folk dancing, as well as different music styles and beat frames. Our work allows greater control of the structure and theme of the dance: we can automatically match a given dance-genre, we can match a specific dance, or even allow manual design: choosing a sequence of specific motifs and filling in the gaps between them to create a coherent dance. 

Our main contributions are summarized as follows:
\begin{itemize}
    \item A framework that generates music-guided dance motion that respects the global structure of the dance genre.
    \item A novel motion perceptual loss to guide the synthesis to follow some high-level motion representations (motion motifs), and control the global culture of the generated motion.
    \item Our method follows the music's content, such as the tempo, to ensure music-to-motion synchronization, and the spectrum, to enforce diversity in the style of the generated movements. 
\end{itemize}

\section{Related Work}
\label{section:Related Work}

\subsection{Data-driven character animation}
Animated motion can be created in two main ways. Either by capturing real human motion and process it, or by manually designing key-frames that are created by animators. However, traditional methods fail to scale to large set of possible motions that might arise in realistic and interacting environments. Aided by the increasing availability of large human-motion capture databases, one way to simulate animation is by traversing through a graph that blend motion clips at connected key-frames~\cite{Kovar:2002,Min:2012}, 
that may additionally support task completion and obstacle avoidance~\cite{Safonova:2007}, or securing style consistency~\cite{Beaudoin:2008,Aristidou:2018:TVCJ}. Additional controls over the animation, as well as the use of linear and kernel methods for blending connected motion clips, have also been investigated to enable interactive and natural character animation~\cite{Lee:2002,Grochow:2004,Treuille:2007,Shum:2008}. 
Data-driven methods though do not encode long-term memory for more contextual synthesis, they are hard to be generalized to large-scale highly dynamic and heterogeneous data (e.g., martial arts or dancing), or to be enhanced with new variations to existing motions.

\subsection{Simulation-based methods}
Simulation-based techniques generate physically plausible animations, taking into consideration the physical constraints on the human skeleton, such as the center of mass for balancing, or the muscular activation~\cite{Yin:2007,Levine:2012:CharacterControl,Hamalianen:2015,Geijtenbeek:2013}, while motion is optimized to complete given tasks~\cite{Clegg:2015,Geijtenbeek:2012}. Indeed, physics-based character animation enables motion synthesis with realistic balancing, motion on uneven terrains, including obstacle avoidance, and allows recovery from falling under extrinsic factors~\cite{Ha:2012,Liu:2005,Liu:2016}. However, the generated motion has been commonly characterized as uncanny, where the absence of simulating secondary movements, or the nuance/style of human motion, makes them look robotic. Moreover, priory and explicitly specifying the objectives, costs, and constraints for each individual task is very challenging, especially for highly stylized and dynamic movements such as dancing. A more recent extension to these works attempts to use less rigid objectives by employing adversarial and reinforcement networks so as to learn those physically-based constraints by examples~\cite{Peng:2017,Lee:2018}.

\subsection{Machine learning approaches} 
Recent developments in deep learning and neural networks have shown promise in controlling articulated characters using convolutional (CNN)~\cite{Holden:2016}, recurrent (RNN)~\cite{Fragkiadaki:2015,Jain:2016}, adversarial~\cite{Wang:2018:ECCV}, or phase-functioned neural networks~\cite{Holden:2017}. These networks have been extensively used in animation for character control using high-level parameters~\cite{Holden:2017,Butepage:2017,Zhang:2018,Pavllo:2018}, or long-term memory~\cite{Martinez:2017,Wang:2021,Gui:2018,Zhou:2018,Wang:2019,Henter:2020} to learn rules and skills for interactive character animation~\cite{Lee:2018}. \AAA{More recently, a number of methods that generate motion variations have been proposed. In some works the generation is controlled by weak control signals ~\cite{Alexanderson:2020,Ling:2020,Ghorbani:2020}, in another work the system generates long-term sequences with user-control~\cite{Habibie:2017}, and there are also methods that learn the interaction of the skeleton with external objects, e.g. a ball~\cite{Starke:2020}, or are driven by other external factors, e.g. speech~\cite{Kucherenko:2019}.} However, current methods provide no intuitive control over the synthesized styles, and they do not encode long memory for more global contextual synthesis. 
Reinforcement learning has also been used for physics based animation to synthesize locomotion tasks such as jumping, running, walking, etc.~\cite{Peng:2017,Peng:2018,Lee:2019}, for learning complex tasks, such as balancing on a ball~\cite{Liu:2017}, or basketball dribbling~\cite{Liu:2018}. In contrast to our method, these frameworks do not generate motion variations, and are limited to a narrow set of use cases.

Despite recent progress in human motion modeling, and deep neural networks that are applied to motion data, motion synthesis is still an open problem. One of the main open challenges is to handle long-term temporal correlations between various motions within a sequence, while preserving a global long term contextual consistency. This challenge becomes even more interesting when attempting to control the generated motion in response to external factors e.g., given music, speech, or the interaction with other characters. This is more evident when dealing with highly dynamic and stylistic movements, such as dancing, a creative and highly expressive form of human movement. Dance choreographies often response to music, including elaborate motor control, rhythm, and synchronization, and reflect to a meaning, an emotion, a story with the help of music and dance moves~\cite{Aristidou:2019}.

\subsection{Audio-driven methods}
Over the years, a number of methods have been introduced to extract the beat rate~\cite{McKinney:2007,Ellis:2007}, or other spectrum-based features~\cite{Mckinney:2003} from audio data, where some are specifically designed for music and motion structure analysis~\cite{Muller:2007}. More recently, machine learning methods have been used to learn audio features~\cite{Suwajanakorn:2017,Taylor:2017,Arandjelovic:2017} that are later used to drive animation, mainly for facial control~\cite{Karras:2017}, speech animation~\cite{Zhou:2018:Visemenet}, or motion synchronization~\cite{Langlois:2014}. 

\subsection{Music-driven character animation}
The intrinsic relationship between dance motion and rhythm motivates us to design and develop a system that generates music-driven character animation that carries the contextual and cultural meaning of dance, what we later call global consistency. Previous works in audio-driven human body animation built movement transition graphs that are synchronized to the beat~\cite{Kim:2003,Sauer:2009,Chiang:2015}, 
or the emotion~\cite{Shiratori:2006, Aristidou:2018:TVCJ}. \AAA{More recent works encode music and motion into a low-dimensional latent space, and then generate dance animation by decoding that latent space, either by using} hidden Markov models~\cite{Ofli:2012}, Gaussian processes~\cite{Fukayama:2014}, recurrent neural networks~\cite{Alemi:2017,Shlizerman:2018,Tang:2018,Kao:2020,Zhuang:2020:Music2Dance}, \AAAA{or autoregressive encoder-decoder networks~\cite{Lee:2019:Chroreography,Ahn:2019}. Others use graph-based frameworks to synthesize dance~\cite{Fan:2012,Lee:2013,Huang:2021}.
The key idea of these approaches is to construct a database of music-motion pairs, and then by finding the best connecting sequences, to compose basic dance movements synchronized to the input music. More recently, 
Chen~\etal~\cite{Chen:2021} introduced ChoreoMaster: the authors created a unified embedding space for music and dance clips, that is fed to a choreography-oriented graph-based motion synthesis framework to generate new dance motions. A similar synthesis-by-analysis learning framework was also implemented using GANs, but only to generate dance in 2D~\cite{Lee:2019:dancing2music}.} Recently, new approaches are focused on adding motion diversity into the synthesis process~\cite{Zhuang:2020,Li:2020}, \AAA{and try to use of more sophisticated architectures, such as transformers~\cite{Li:2020,Li:2021}}. However, all of these methods do not take into consideration the global content of the dance movement, generating motions that indeed follow the given audio beat, with or without style variation, but are contextually meaningless, while, in some cases, the generated motions are constantly repeated. To the best of our knowledge, our work proposes the first generative model that enables generation of diverse motions where the global content of the choreography remains consistent. The choreography has only been considered as a high level representation of the moves of multiple characters in the space~\cite{Schulz:2011,Soga:2016,Iwamoto:2018} rather than generating a new dance variation, driven by an input music. \AAAA{In a similar concept to our hierarchical dance composition, Ye~\etal~\cite{Ye:2020} has recently proposed ChoreoNet - a two-stage music-to-dance synthesis framework. However, while their approach is two-stage, bottom-to-top approach, which includes: (1) generating independent small motion units and (2) stitching them together through (inpainting), it does not takes into consideration the global choreographic structure; our approach is 3-level top-to-bottom, where the motion generation is driven by the (1) high-level signature (choreography) which dictates the order of the (2) motifs, which drives the generation of individual poses (3).}


Zhou~\etal~\cite{Zhou:2018} proposed a generative network that outputs long term smooth motion sequences, learned from motions of kinematic skeletons. Our work is built upon their core model that uses a recurrent network which simply receives and outputs joint positions of skeleton poses. However, our adaptation contains a few major differences: (a) we use rotational rather than positional data to avoid pose ambiguities that is critical for highly dynamic and complex movements, such as in dancing; (b) our extended version is additionally conditioned by 3 other factors, including audio features (4D), the motif representation (184D), and the foot contact labels (2D); (c) the addition of a motion perceptual loss, to enable control over the choreography of the generated motion.


Aristidou~\etal~\cite{Aristidou:2018}, presented the concept of \emph{motion signatures} as a descriptive and temporal-order invariant representation of motion sequences. To create a signature a sequence is broken down into small overlapping movements, named \emph{motion words} and the signature is a collection of some embedded representation of these words, called \emph{motifs}. While their work is focused on motion analysis, where the content of motion can be distinguished using learned semantic features, our work adapts this premise for motion synthesis. The use of motion motifs as a prior to the generation of motion words and the fact that it enables control over the dance synthesis to respect the global structure of choreography, is original.


\section{Data Representation}
\label{section:Data Representation}


Dance motion is highly correlated to music~\cite{Muller:2007}. Hence, our goal is to generate a motion sequence conditioned by the input music. To achieve this, we need to extract acoustic and motion features, and define a mapping between them.
In this section we present details about the audio and motion features used in our framework. In particular, we describe the rhythm and spectral features extracted from the input music clip, and discuss the representation of motion entities in the 3 different levels: pose, motif and choreography.

\subsection{Audio Representation}
\label{subsection:Audio Representation}

In music theory, various factors should be considered in order to faithfully represent a music signal. Similarly to~\cite{Alemi:2017}, our audio representation consists of several features that can be divided into two main groups: \emph{rhythmic features} that describe the rhythm and the tempo of the music, and \emph{spectral features} that are obtained by mapping the audio signal to the frequency domain commonly used to identify the notes, pitch, and melody of audio streams. 



\subsubsection{Rhythmic features}
\label{subsubsection:Beat tracking}
Rhythm is the element of time in music. In general, rhythm is characterized by the \emph{beat}, that is the regular pulsation; the \emph{tempo}, that is the rate of speed or pace of the musical pulse; and the \emph{meter}, that is the grouping of beats into larger, regular patterns, notated as measures. To extract these rhythmic features we employed the well-known Librosa Library~\cite{Ellis:2007,McFee:2015}; rhythmic features are computed per frame using a temporally moving right-anchored window (10 frames) with step one frame. We use four acoustic features to encode the rhythm $\mathbf{a}_r^t = [a_1,\dots,a_4]$, as listed in Table~\ref{tab:audio features}, in the Appendix. 

Special attention is given to beat detection since it is used to divide the motion sequences into meaningful temporal kinematic entities that are later used to drive the animation. \AAAA{We represent beats as temporal binary signals, and the process of their detection uses the analysis of audio signals, onset detection, and the study of repetitions to determine which onsets constitutes the actual beats~\cite{Bello:2005}.} 

\subsubsection{Spectral features}
\label{subsubsection:Audio features}

The spectral features are obtained by mapping the audio signal to the frequency domain using a Fourier Transform. These features are commonly used to identify the notes, pitch, and melody of audio streams. Again, we consult the Librosa Library to extract those spectrum-related features, which are described in Table~\ref{tab:audio features} in the Appendix. Spectral features are sampled per window that is formed between two consecutive audio beats. The spectral features on every beat $k$ are denoted by $\mathbf{a}_s^k \in \mathbb{R}^{87}$.

\AAA{Note that there are other spectral representations of audio that can be incorporated into the proposed network, such as the Short-time Fourier transform (STFT), or the Continuous Wavelet Transform (CWT) features. Similarly to~\cite{Li:2020,Tang:2018}, we only use the features integrated in the Librosa library as we noticed that they add enough variation to the outputs, and they also help the acLSTM network to be synchronized with the beat.}

\subsection{Pose Representation}
\label{subsection:Motion Representation}

A skeleton pose is represented as a combination of $J=31$ joint rotations, in a hierarchical order, using unit quaternions, $\mathbf{f}_{q} \in \mathbb{R}^{4 J}$, together with the global root 3D position. The position of the root at time $t$, $\mathbf{f}_{r}^{t} \in \mathbb{R}^{3}$, is represented as the displacement from its position in the previous frame. This representation works well with cycling and repetitive motions, and allows smooth motion synthesis. Therefore, at time $t$, a skeleton pose is defined as $\mathbf{f}^{t} = [\mathbf{f}_{r}^{t},\mathbf{f}_{q}] \in \mathbb{R}^{3 + 4 J}$. \AAAA{Since quaternions may exhibit temporal discontinuities due to the antipodal property ($\mathbf{{f}_{q}}$ and $-\mathbf{{f}_{q}}$ represent the same rigid transformation), we adapt the technique proposed by~\cite{Pavllo:2018}, where we choose among $\mathbf{{f}_{q}}$ and $-\mathbf{{f}_{q}}$ the representation with the lowest Euclidean distance from the representation in the previous frame.}

\subsection{Motif Representation}
\label{subsection:Motif Representation}

We define a \emph{motion word} as a collection of $N$ consecutive skeleton poses, representing the local evolution of pose. More specifically, motion words are narrow temporal-windows of skeletal poses, which divide a motion sequence into smaller feature descriptors, and can be expressed as $w \in \mathbb{R}^{N \times |\mathbf{f}|}$~\cite{Aristidou:2018}.  

The kinematic beat is usually synchronized to the music beat pattern~\cite{Davis:2018}, so that small temporal motion events usually start and end on the beat. Similarly, in our implementation, motion sequences are divided into motion words on the beat. The size of motion words depends on the beat rate of the music, and there is no overlap between two consecutive motion words. Thus, each motion word represents a short but meaningful movement. 
\begin{figure}[t]
	\includegraphics[width=\linewidth]{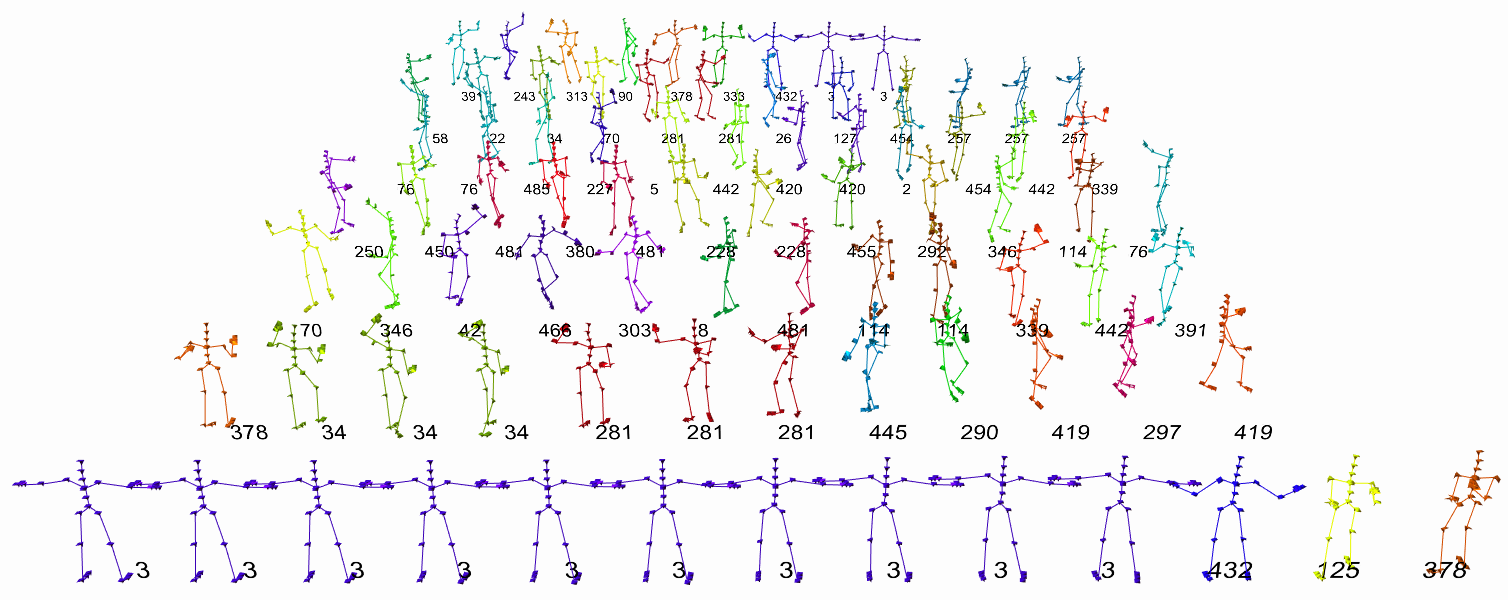}
	\caption{Motion words and motifs. A whole motion sequence (displayed from bottom left to top right) is divided into short temporal segments, called motion words (here represented by one pose). Each word is clustered in the embedding space and associated with a motion motif, which is the  the centroid of the cluster (here represented by the color of the word and ID number).}
	\label{fig:motion_words}
\end{figure}

Given a group of similar motion sequences from our dataset $\mathcal{D}$ (e.g.\ from one type of dance), we map all motions words extracted from the group into a $d$-dimensional universal feature space $\mathbb{R}^d$ similar to~\cite{Aristidou:2018}, but use a smaller dimension $d=184$ to create a more compact representation for the embedding. \AAA{In contrast to the original paper where motion words have a fixed number of frames, in our case they have different length due to the temporally-variant beat rate. Thus, given the longest word, we time-scale the motion words to have a uniform size of 13 frames (which is 140bpm @30fps).} We then cluster the motion words using K-means clustering into $K=500$ clusters. We have found that segmenting dance motion sequences on the beat assists in creating more compact and better separated motion clusters. Each cluster $\mathcal{M}_i$, is represented by one \emph{motif} word, $\mathbf{m}_i\in\mathbb{R}^d$, which is chosen to be the motion word that is closest to the centroid of the cluster, using $\ell_2$ distance. Motion motifs are small movements that are common inside a motion sequence and reveal its content and characteristics. Any motion word can now be associated to a cluster simply by finding its closest motif. Figure~\ref{fig:motion_words} visualizes a motion sequence that is divided into motion words, where each is associated with a different motif (indicated by a color and a cluster id).



\subsection{Choreography Representation}
\label{subsection:Choreography Representation}

\begin{figure}[t]
	\includegraphics[width=0.49\linewidth]{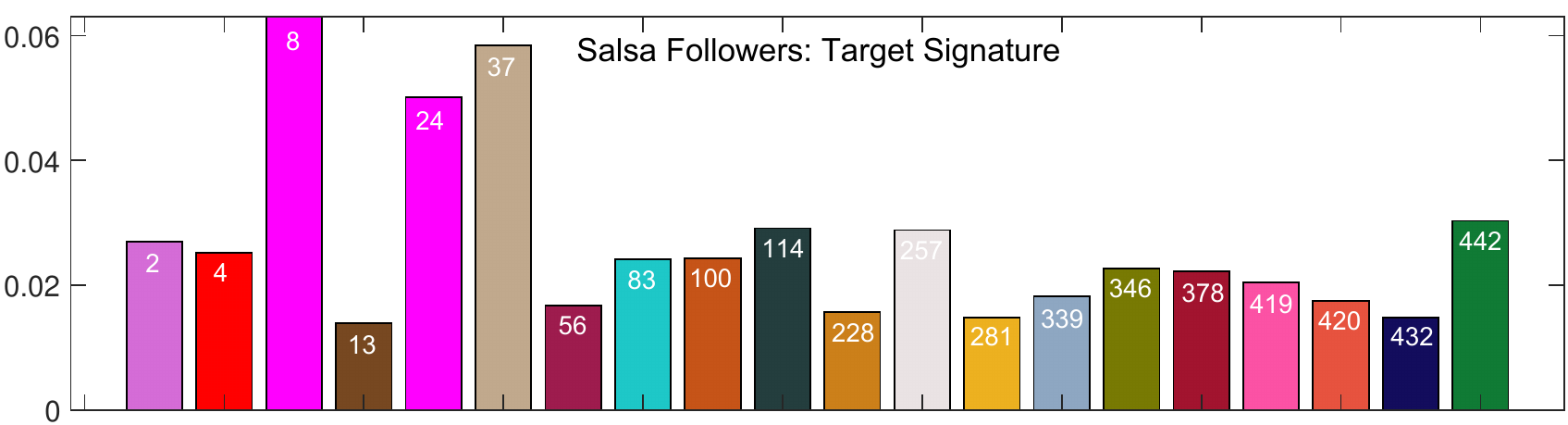}
	\includegraphics[width=0.49\linewidth]{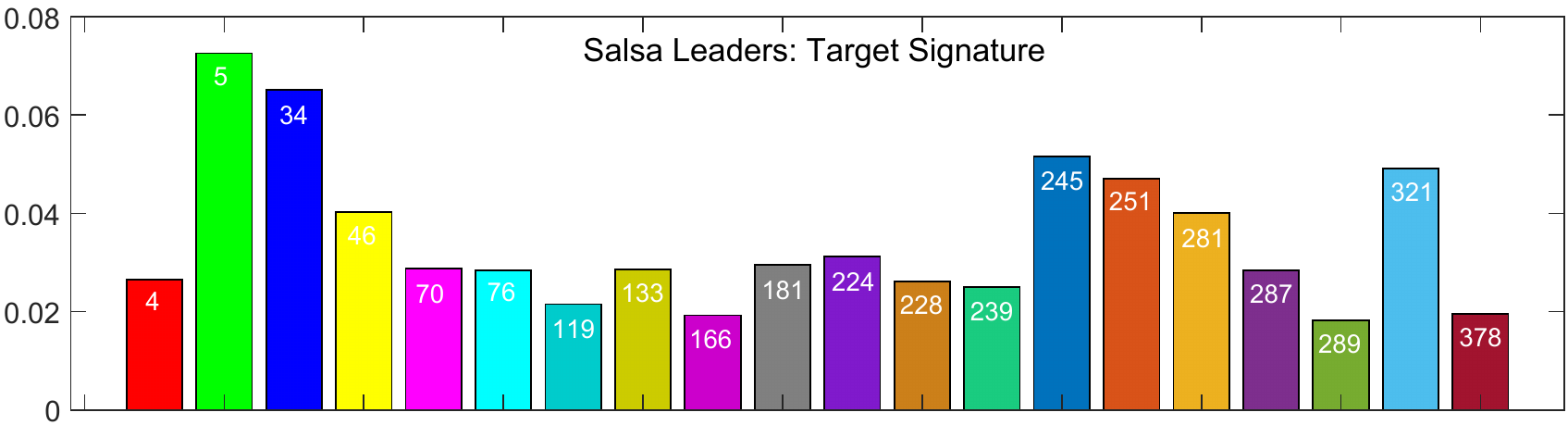}
	\includegraphics[width=0.49\linewidth]{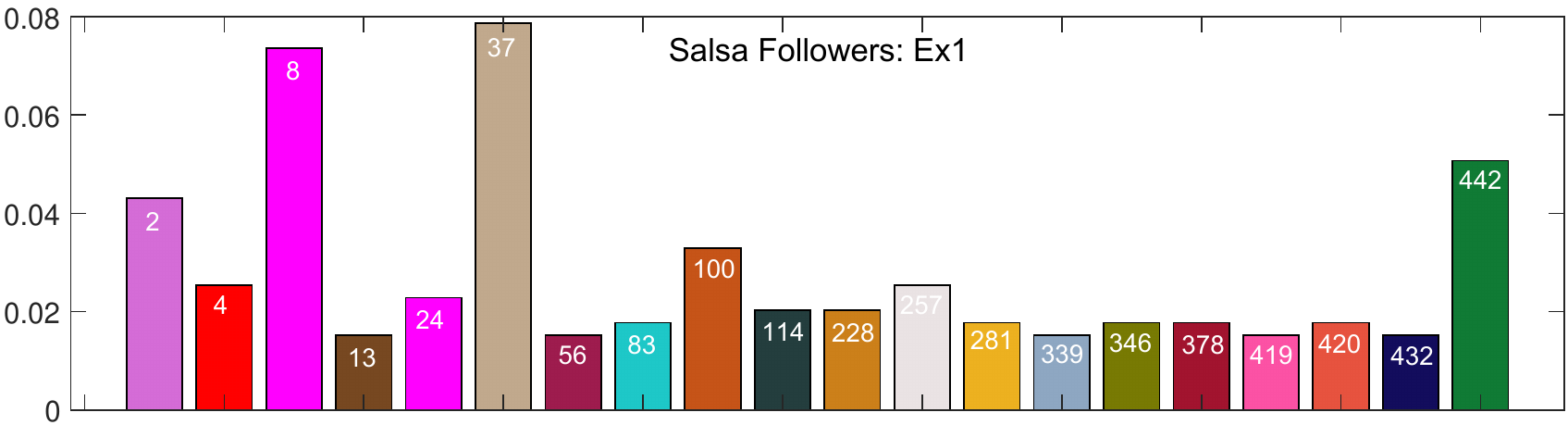}
	\includegraphics[width=0.49\linewidth]{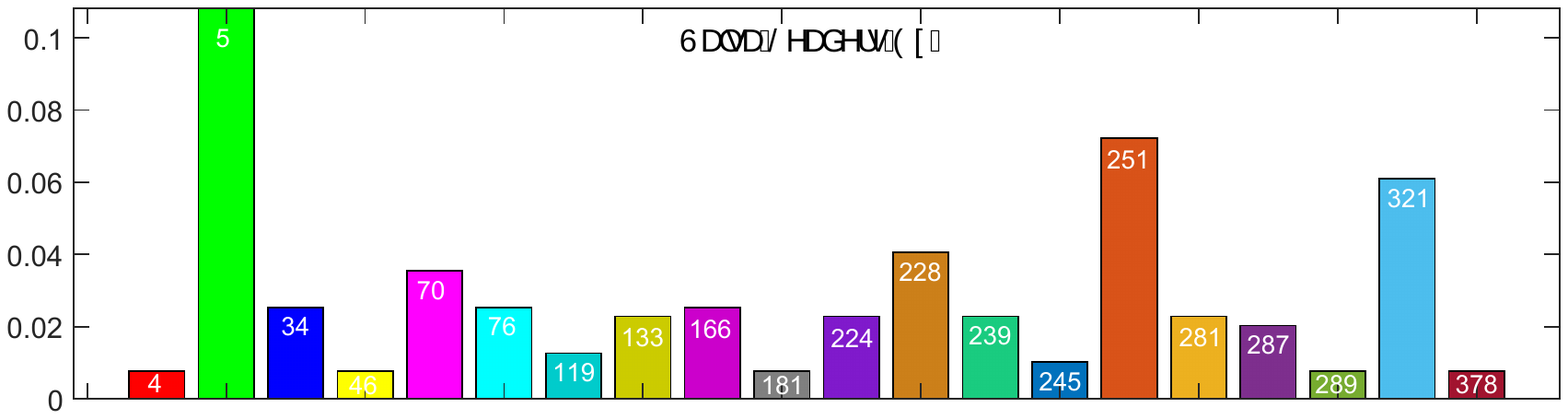}
	\includegraphics[width=0.49\linewidth]{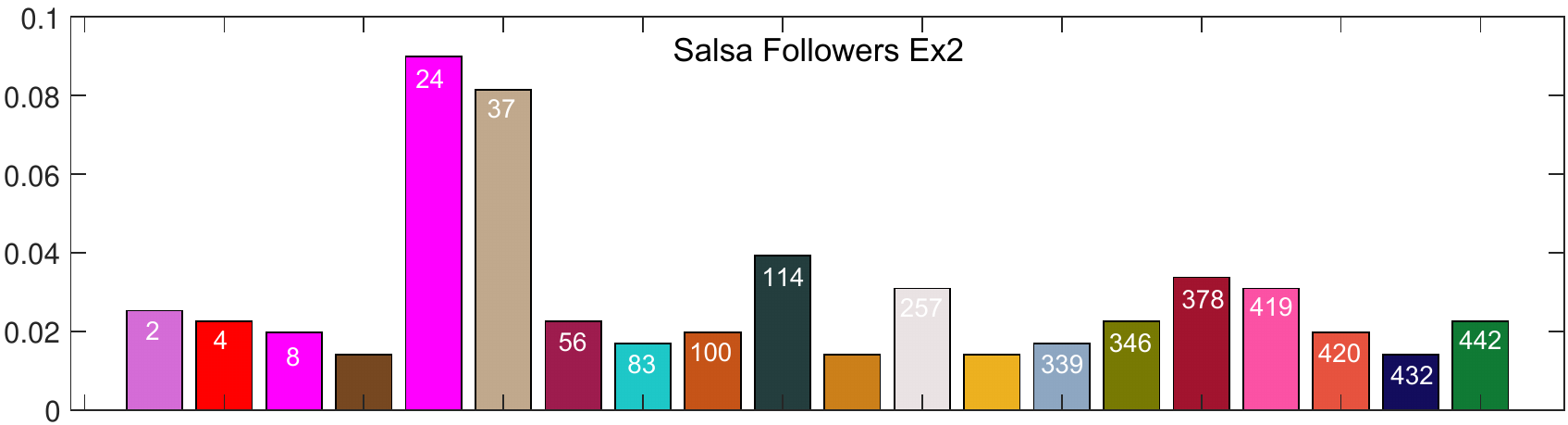}
	\includegraphics[width=0.49\linewidth]{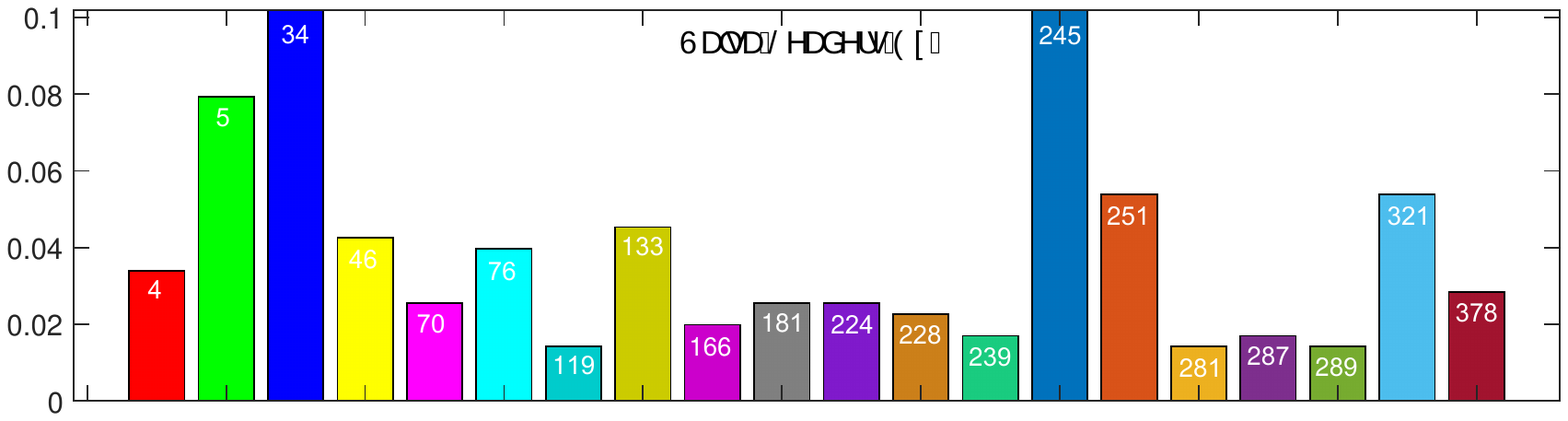}
	\includegraphics[width=0.49\linewidth]{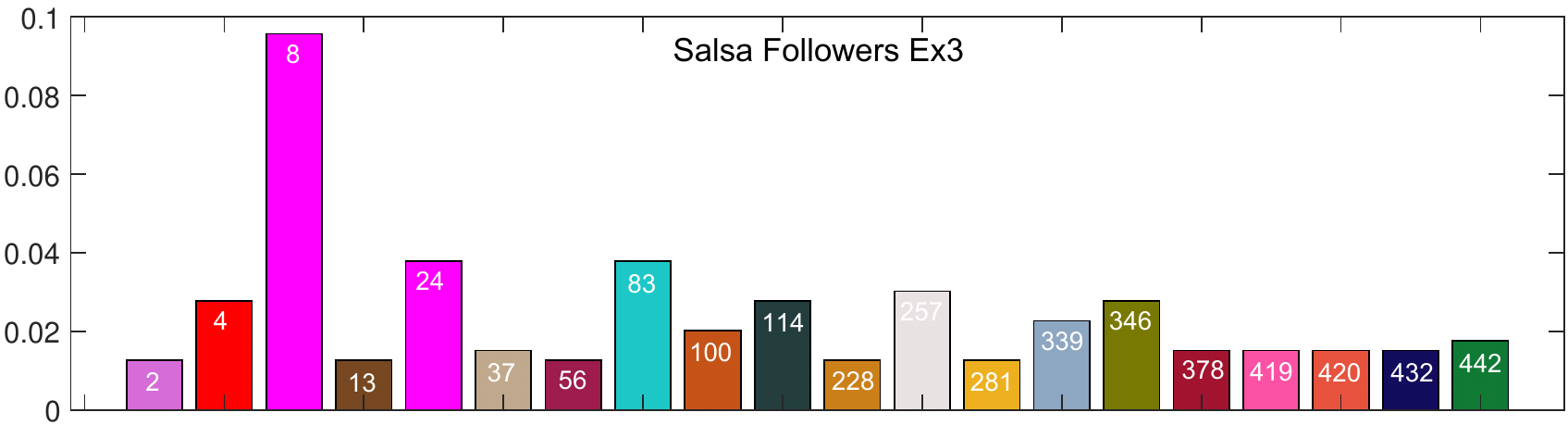}
	\includegraphics[width=0.49\linewidth]{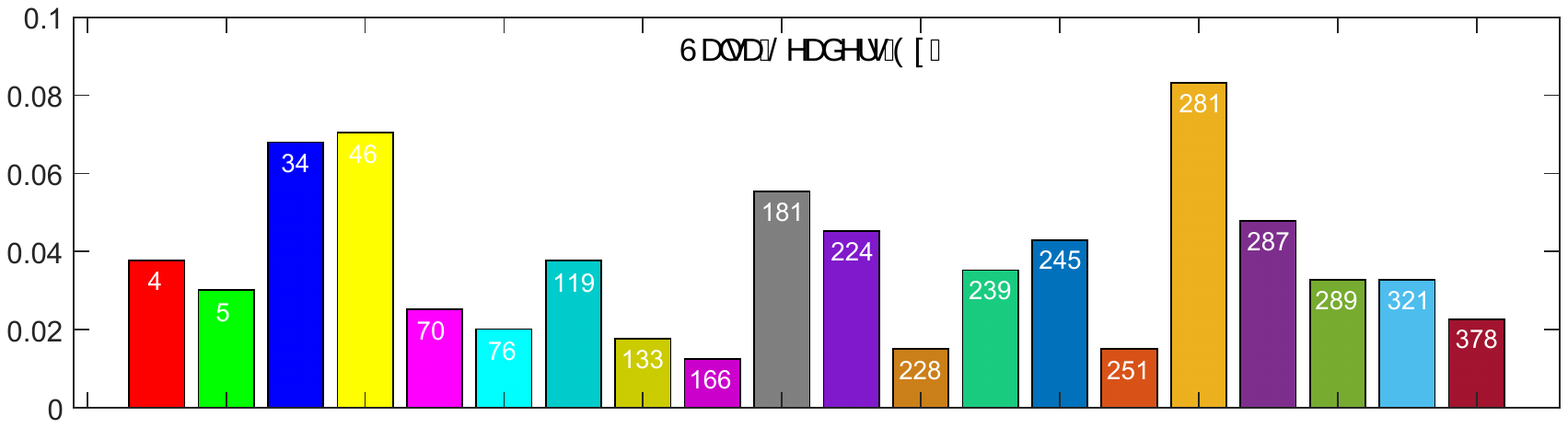}
	\caption{Motion Signatures: the bars' color indicate the cluster id and the height shows the frequency of the motion word in the motion sequence. We only illustrate the distribution of the 20 most frequent words. The right column shows motion signatures of a salsa leader and left column of a salsa follower. The three bottom rows are from specific dances and the top row is the template motion signature we built for this genre.}
	\label{fig:Signatures}
\end{figure}

A dance does not only reflect the synchronization of local motions to the music, but also carries a specific structure and set of movements. At the choreography level, we represent each dance as a bag-of-motifs. We extract the \emph{motion signature}, which is the distribution of the motion motifs of the dance. This signature provides a succinct but descriptive representation of motions sequences that is semantically meaningful, it is time-scale invariant and agnostic to the motif order. Such signatures carry critical global information about the content of a dance sequence as a function of motifs. Figure~\ref{fig:Signatures} shows motion signatures of specific dances of a salsa leader and follower. Note that the dances share many motion words, but with different frequency of appearance. We create a combined signature templates (shown at the top) for a specific genre by averaging many dances.
Such signatures at the choreography level will later drive our synthesis process of the dance.



\section{Music-driven Dance Generation}
\label{section:Music-to-Dance Generation}

\begin{figure*}[t]
	\includegraphics[width=\linewidth]{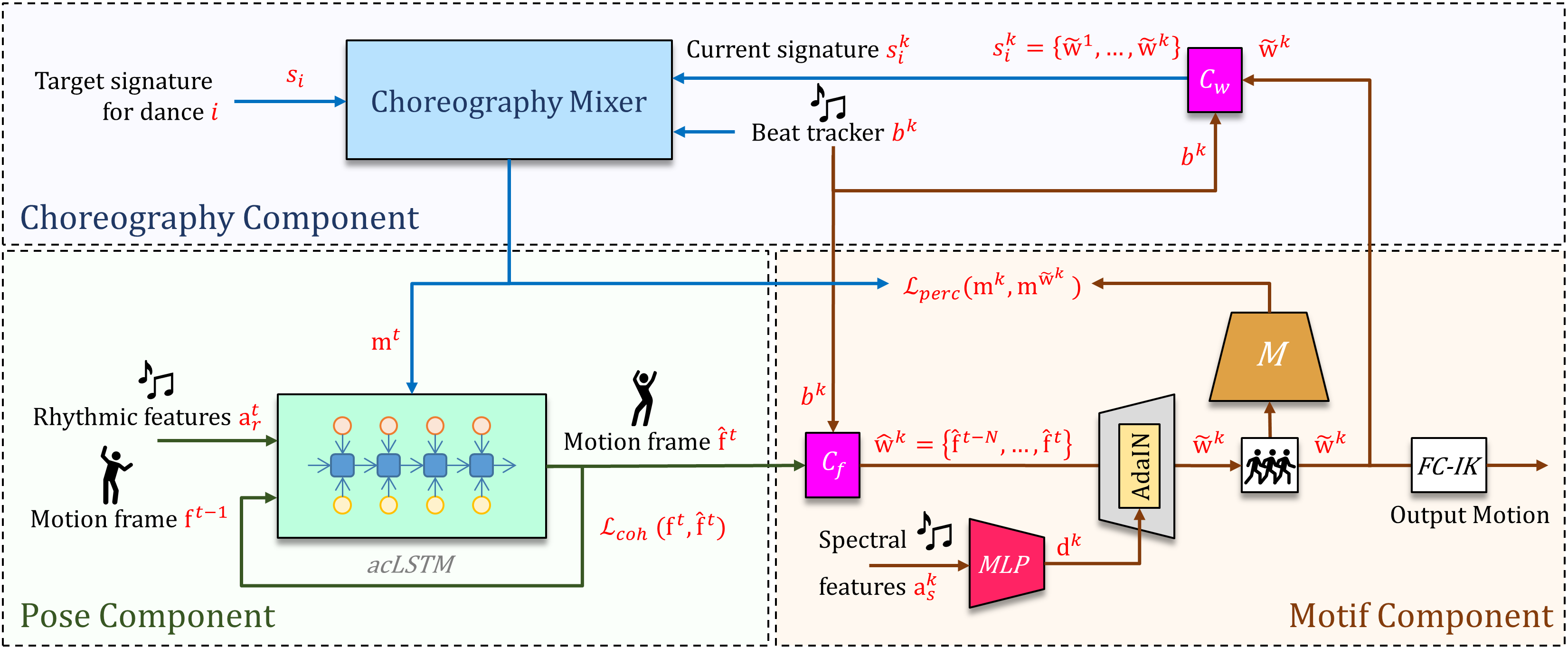}
	\caption{Our framework consists of three main components: pose, motif and choreography, where each is responsible for a different resolution in the motion synthesis process. The pose level generates a sequence of temporal coherent poses; the motif level ensures that a sequence of poses creates a short movement that follows a specific motion motif; the choreography level takes care of the global, long-term, structure of the  dance by generating the sequence of motifs.}
	\label{fig:Network_Overview}
\end{figure*}

Our framework consists of three main levels, in a hierarchical order: (a) the \emph{pose} level (per frame), that learns how to create temporal coherent moves according to the rhythm, (b) the \emph{motif} level (per motion word), that learns how to perform short natural movements with some variation that are typical to the dance style and are controlled by the input audio features, and (c) the \emph{choreography} level (per motion signature), that learns the higher level distribution of dance movements within a dance genre, and generates meaningful dance movements with global consistency. A detailed description of our framework is illustrated in Figure~\ref{fig:Network_Overview}.

\subsection{Learning to Move}
\label{subsection:Learning How to Move}


The purpose of the pose level is to generate a sequence of skeleton poses $\mathbf{f}^t$ that are long-term stable and temporally coherent. Thus, we adapt the auto-conditioned long short-term memory (acLSTM) network of Zhou~\etal~\cite{Zhou:2018} \AAA{that consists of three fully connected layers}. We denote this network by $R$. This network learns the temporal evolution of dance and deals with nonlinear dynamics. To cope with long-term predictions the network is fed by its own outputs during training, which enables generation of stable sequences in test time. 

While in the original work, acLSTM is a recurrent network that simply receives and outputs skeleton poses, which are represented by joint positions, our adaptation contains a few major differences. First, our network works with joint rotations, which enables us to exploit and learn the rotational information existing in the dataset, and to directly extract high-quality animation files. In contrast, in the original work, the acLSTM generates joint positions, a representation which cannot be used to animate a 3D characters, may lead to temporally inconsistent bone-lengths, and requires an additional post processing step to infer rotations from positions. 

Another major difference is that in our case, our LSTM $R$ is conditioned by 3 other factors:
\begin{enumerate}
    \item Audio rhythmic features $\mathbf{a}_r^t \in\mathbb{R}^{4}$ -  which assist the network by indicating in advance how long it will take to complete the creation of the next motion word, when the length of the motion word is determined by the gap between two consecutive beats.
    \item A motif $\mathbf{m}^t\in\mathbb{R}^{184}$ -  which guides the network to generate a sequence of poses that belong to a specific cluster of motions, enabling high-level control over the movements displayed in the choreography level. 
    \item Foot contact labels $\mathbf{c}^t\in\{0,1\}^{2}$ - a binary vector representing the left and the right foot contact labels (`1'-- if the foot is in contact with the ground, '0' -- if not). These labels are extracted during pre-processing, by examining the distance of the foot joints from the ground and their velocity.
\end{enumerate}







Therefore, the input to the network at time $t$ is the vector \(\mathbf{n}^t\) which includes the motion vector, concatenated with the audio features, the motif vector, and the foot contact labels:
\begin{equation}
    \mathbf{n}^t = [\mathbf{a}_r^t, \mathbf{m}^t, \mathbf{f}^t, \mathbf{c}^t],
\end{equation} 
and the output of the network is the next predicted skeleton pose and foot contact labels, given by
\begin{equation}
     [\hat{\mathbf{f}}^{t+1},\hat{\mathbf{c}}^{t+1}] = R(\mathbf{n}^t) 
\end{equation} 

\textbf{Foot Sliding Cleaning:} In general, recurrent generative models synthesize motions that suffer from artifacts such as foot skating, mainly because their prediction which is fed back into the input accumulates errors, both in time and along the kinematic chain. To avoid those common artifacts, we train the network to predict foot contact labels. This step by itself, with no further processing, improves the quality of our outputs. However, to completely remove foot sliding artifacts, we also use standard Inverse Kinematic (IK) optimization, on corrected foot positions, as a post-processing step. 
%

\subsection{Learning to Dance}
\label{subsection:Learning How to Dance}


In the next level of the hierarchy, we present the \emph{motif} level whose purpose is to collect a temporal set of poses, which is equivalent to a motion word, and ensure that it belongs to the distribution of a specific cluster, enabling high-level control of  movements in the dance. 


In practice, on every beat $k$, a temporal sequence of $N_k$ generated poses is collected ($N_k$ may vary between pairs of consecutive beats) to form one motion word $\hat{\mathbf{w}}_k = \{\hat{\mathbf{f}}_{t_{k-N_k-1}},\ldots,\hat{\mathbf{f}}_{t_k}\}$, where $t_k$ is the index of the frame that corresponds to the $k$th beat.
Thus, the motif level can be described as a higher-level network $H$ whose input is the current motion word together with the next motif and the audio feature, and outputs the next motion word by
\begin{equation}
\left[\hat{\mathbf{w}}^{k+1},\hat{\mathbf{c}}_{w}^{k+1}\right] = H\left(\mathbf{w}^{k},\mathbf{c}_{w}^{k}, \mathbf{m}^{k+1}, \mathbf{a}^{k+1}\right),
    \label{eq:motiflevel}
\end{equation}
where $H(\cdot)$ is $N_k$ compositions of $R(\cdot)$, $\mathbf{c}_{w}^{k}$ is the concatenated foot contact labels that correspond to $\mathbf{w}^{k}$, and $\mathbf{m}^{k+1}$ and $ \mathbf{a}^{k+1}$ remain constant per motion word. 

To ensure that the collected group of poses $\mathbf{w}^{k+1}$ exhibit a certain movement as a motion word, we next propose a motion perceptual-loss, that is utilized during training.

\subsubsection{Motion Diversity}
To convey diversity and richness in the synthesized motions, we enable subtle variation in the generated motion words. This variation is controlled by the spectral audio features, $\mathbf{a}_s^t$, that are responsible for the style of motion~\cite{Yang:2012}. In this manner, we add an Adaptive Instance Normalization (AdaIN) layer that applies a set of temporally invariant affine transforms to the motion words, where the joints are treated as channels. The AdaIN parameters are extracted by a Multi-Layer Perceptron (MLP) component that receives the spectral features $\mathbf{a}_s^k \in \mathbb{R}^{87}$, and outputs a normalized vector of 248 parameters (2 times the size of $\mathbf{f}_{q}\in \mathbb{R}^{124}$) that is tiled along the temporal axis resulting in $\mathbf{d}^{k} \in \mathbb{R}^{248\times N}$. Then the AdaIN layer uses these parameters to apply a scale and translation for each joint rotation in the motion word. 

In order to maintain the temporal coherency between consecutive motion words, we multiply the tiled parameters, by a temporal weighting function, $\gamma(t)\in \mathbb{R}^N$, in the form of an Hamming window that is centred in the middle frame of the motion word, and decays towards its edges (the first and last frames of the motion word remain unmodified). The modified motion word is given by
\begin{equation}
\tilde{\mathbf{w}}^{k} = \left(1 + \beta \times\gamma(t) \times \mathbf{d}_{(1)}^{k}\right) \hat{\mathbf{w}}^{k} + \beta \times \gamma(t) \times \mathbf{d}_{(2)}^{k},
\label{eq:diversity}
\end{equation}
where $\beta$ is a scalar that controls the amount of variation we apply (in our experiments, we used $\beta = 40$), and ${d}_{(1)}^{k}\in\mathbb{R}^{124\times N}$ and ${d}_{(2)}^{k}\in\mathbb{R}^{124\times N}$ are the scale and translation parameters extracted from $\mathbf{d}^{k}$, respectively. 

Since the parameters are temporally invariant, the shape of the individual signals (joint rotations), and the motion content (motifs), is preserved but at the same time, it enables the synthesis of different motion variants that are correlated with the spectral features. In order to guarantee that the resulting motion belongs to the driving motif, we incorporate a motion perceptual loss.

\subsubsection{Motion Perceptual-Loss}
In the image domain, perceptual-loss functions are used to define high-level differences between two images by comparing their semantic representations~\cite{Johnson:2016}. To extract such a representation for images, previous works commonly use pre-trained classification networks that were trained on million of images, such as VGG-16~\cite{Simonyan:2015}, and use their deep features as a representation for semantic information. 

We propose a similar notion of \emph{motion perceptual-loss}. Our motivation is to have a component that observes the content of a group of frames and guides the corresponding movements to belong to a specific cluster of motions. However, in contrast to images, there are no large-scale pre-trained motion classification networks, and it is challenging to semantically label a dataset of short movements to train a motion classification network. Instead, we use a pre-trained network proposed by Aristidou~\etal~\cite{Aristidou:2018}, which maps low-level motion words to an embedding space that constitutes a high level representation of motion. While training $P$, similar motions are encouraged to map into the same region in the embedding space. The network is trained to distinguish between groups of similar motions disregarding low level details, temporal warping and frame rate differences. \AAAA{Note that, the term ``pose perceptual loss'' was also used by Ren~\etal~\cite{Ren:2020}, which use an activity recognition network to extract features from 2D pose data. In our case, we use the perceptual loss in the context of 3D motions, and our features are extracted from a network that tries to map similar motion words into nearby points in the embedding space. The latter enables the extracted features to contain information related to the spatio-temporal nature of the movement itself (e.g., ``right arm is raised up''), while such information will be ``cleaned'' by deep features extracted from an activity recognition that tries to recognize the high-level semantic meaning of the motion (e.g.,``waving hands'').}

\subsection{Learning the Choreography}
\label{subsection:Learning the Choreography}


The \textit{choreography} level is responsible to maintain the global structure of a dance genre, by controlling the distribution of motifs in the bag-of-motifs representation. The choreography component ensures that the distribution of generated motion words matches, in long-term, the distribution of the motion words for the given dance genre, maintaining globally the culture of that dance. The distribution of generated motion words, a.k.a.\ motion signature, is represented by a normalized histogram $\mathbf{s}\in\mathbb[0,1]^K$. 

More specifically, the choreography component chooses the next motif in the dance based on two criteria:
\begin{enumerate}
    \item Motif Transition - a matrix $T$ that describes probability of the temporal connectivity between consecutive motion motifs for a given dance genre is used. The matrix is computed by counting appearances of each pair in real dances in the genre. This enables considering the transition probability between different motifs. 
    \item Signature difference - the current calculated signature $\mathbf{s}^k$ is compared to the target signature of the dance genre $\mathbf{s}^G$, and guides the network to generate missing motifs by raising the probability of these motifs.
\end{enumerate}

In practice, on every beat, the selection is performed by randomly drawing the index of the next motif ($k+1$) from a probability distribution that is calculated by 
\begin{equation}
    \mathbf{p}_j^{k+1} = |\mathbf{s}^k - \mathbf{s}^G|\cdot T_j
\end{equation}
where $T_j\in\mathbb{R}^K$ is the transition probability vector from the $j$th motif to the other motifs. The selected motif $\mathbf{m}^{k+1}$ is then set as ``target'' input to the acLSTM network, to generate motion that corresponds to the cluster of the chosen motif.

\subsection{Training and Loss}
\label{subsection:Training and Loss}
Our dataset, which is described in detail in Section~\ref{section:Results and Discussion} is divided into short motion words, based on the music beat, which comprise our motion collection $\mathcal{D}$. Training is performed using sequences of 100 consecutive frames for 500000 iterations using the Adam back propagation algorithm~\cite{Kingma:2015}. The initial learning rate is set to 0.0001 and the batch size to 32. We next define the losses applied to the various components within our pipeline.

\subsubsection{Temporal Coherence Loss}
\label{subsubsection:Temporal Coherence Loss}
At pose level, 
%
we employ the pose loss of~\cite{Lee:2002}  for our objective function. This loss uses a weighted sum of the difference of rotation between joints, which is given by
\begin{equation}
d_q(\mathbf{f^j}, \mathbf{f^i}) = \sum_{l=1}^{J} \parallel \log \left(q_{j,l}^{-1} q_{i,l} \right)\parallel^2,
\label{eq:Distance_Metric}
\end{equation}
where $J$ is the number of joints in the motion word, and $q_{i,l}$, $q_{j,l} \in \mathbb{S}^3$ are the complex forms of the quaternion for the $l$-th joint in the $i$ and $j$ frames, respectively. The log-norm term $\parallel \log \left( \cdot  \right)\parallel^2$ represents the geodesic norm in quaternion space, which yields the distance from $q_{i,l}$ to $q_{j,l}$ on $\mathbb{S}^3$. In addition, to mitigate error accumulation of joint rotations along the kinematic chain, we added a forward kinematic layer that is applied to the output rotations~\cite{Pavllo:2018}. This allows us to compare the resulting positions to the ground truth positions through an additional loss on positional data.

We train our acLSTM network, using both ground truth pose frames and fake output pose frames. If at time $t$ the network $R$ receives the ground truth, the loss is given by:
\begin{equation}
\label{eq:coh}
  \mathcal{L}_{coh} = d_q\left( R\left(\mathbf{f}^t, \mathbf{m}^t\right),  \mathbf{f}^{t+1}\right),
\end{equation}
where $\mathbf{f}^t$ and $\mathbf{f}^{t+1}$ are two consecutive poses extracted from the sample data, and $m^t$ is the ground truth motif. Alternatively, if at time $t$ the network receives its own previous output pose, the loss is given by
\begin{equation}
\label{eq:coh_ac}
 \mathcal{L}_{ac\_coh} = d_q\left( R\left(\hat{\mathbf{f}}^t, \mathbf{m}^t\right),  \mathbf{f}^{t+1}\right),
\end{equation}
where $\hat{\mathbf{f}}_t = R\left(R\left(\dots R\left(\mathbf{f}^{t-n}, \mathbf{m}^{t-n}\right), \mathbf{m}^{t-1}\right), \mathbf{m}^t\right)$, and $n$ indicates how many times the network has fed itself in its own input since the last injection of ground-truth pose. Note that the foot contact labels and audio features were omitted in \eqref{eq:coh}, \eqref{eq:coh_ac} and \eqref{eq:percp} for ease of notation.

\subsubsection{Motion Perceptual Loss}
\label{subsubsection:Content Consistency Loss}
The motion perceptual loss is computed by the difference between the embedding of the output motion word and the embedding of the motif ground truth, and given by
\begin{equation}
\label{eq:percp}
  \mathcal{L}_{perc} = \Vert P(H(\mathbf{w}^{k}, \mathbf{m}^{k+1})) - \mathbf{m}^{k+1} \Vert^2,
\end{equation}
where $P$ is the pre-trained network that was trained to map short movements into high-level embedding space, as described in the previous section.

The full loss for our generative motion synthesis model, which is a combination of the two temporal coherence terms, and the perceptual loss, is given by
\begin{equation}
\label{eq:Final Loss}
  \mathcal{L} = \mathcal{L}_{coh} + \mathcal{L}_{ac\_coh} + \lambda\mathcal{L}_{perc},
\end{equation}
where in our experiments $\lambda = 0.5$.

\section{Results and Discussion}
\label{section:Results and Discussion}

In this section, we present the implementation details, and the dataset used for training and testing our method. We also provide several experiments to demonstrate the efficiency of our framework, and conducted extensive studies to evaluate its performance in terms of realism, music synchronization, multi-modality, and global content consistency. 

\begin{figure}[t]
	\includegraphics[width=1\linewidth]{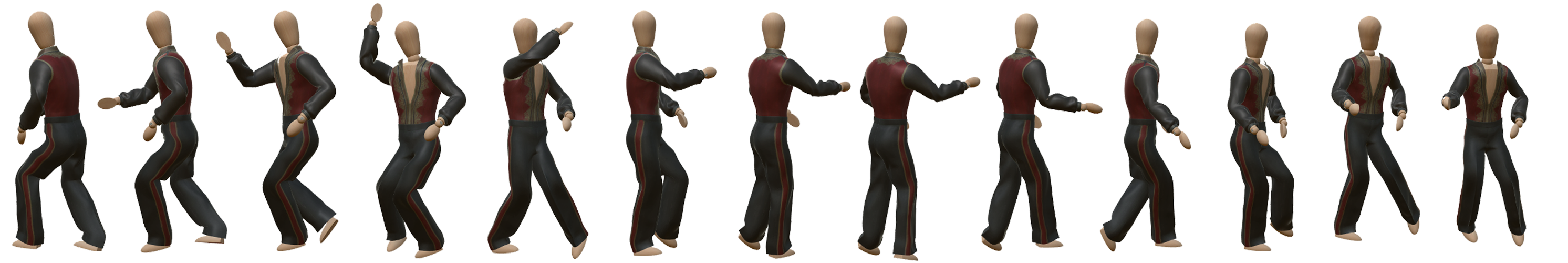}
	\includegraphics[width=1\linewidth]{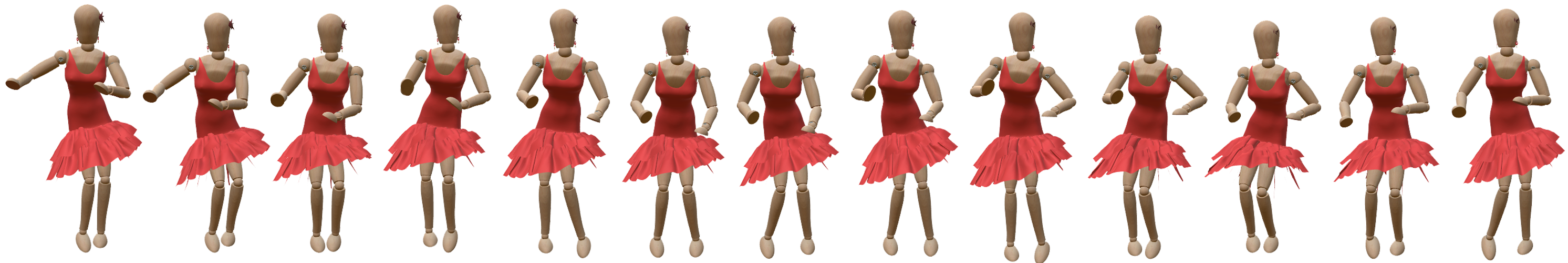}
	\includegraphics[width=1\linewidth]{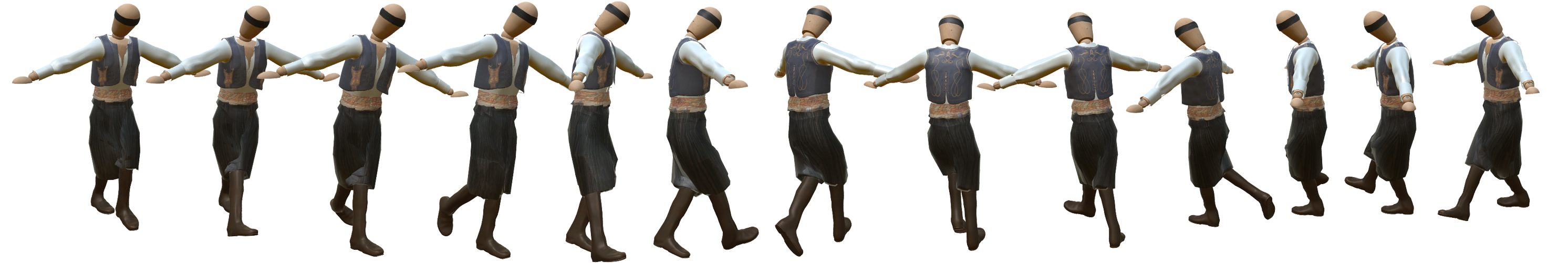}
	\includegraphics[width=1\linewidth]{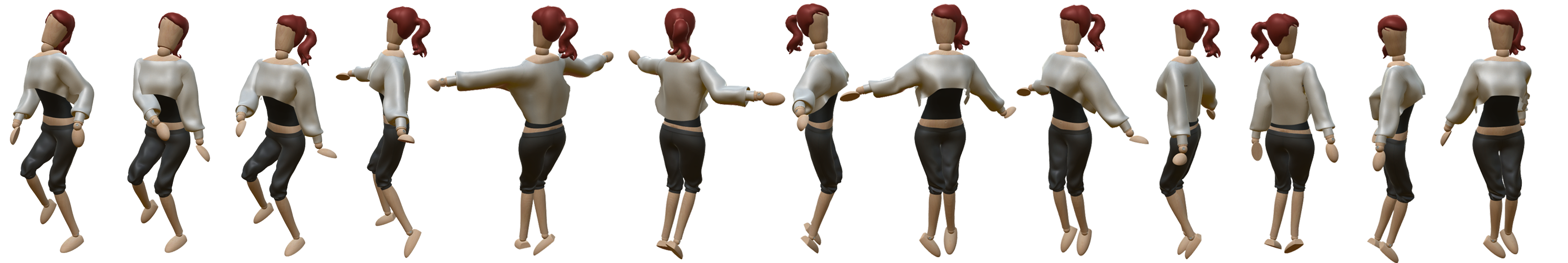}
	\includegraphics[width=1\linewidth]{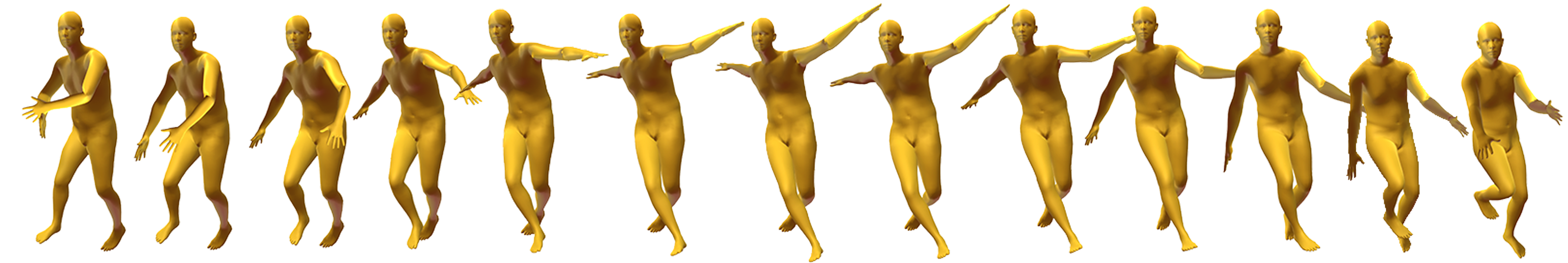}
	\caption{A Gallery of selected frames extracted from our results for various dance genres. Top to bottom: salsa leader and salsa follower, Greek folk dance, \AAAA{modern dance, and hip-hop (from the AIST++ database).}}
	\label{fig:DanceResults}
\end{figure}

Figure~\ref{fig:DanceResults} shows a gallery of selected frames extracted from our results using various dance genres: salsa leader and follower (first and second row, respectively), Greek folk dance (third row), and modern dance (bottom). \AAAA{Note that a different instance of our network has been trained for each dance genre.} To produce the animations, we convert the tensors produced by our network into BVH files, apply a smoothing filter to the global orientations, and use Autodesk Maya (software) to apply skinning to the extracted skeletons. The quality of the resulting animations may be examined in the supplementary video~\footnote{Unfortunately, due to the massive spread of the Covid-19 virus, and the restrictions from the authorities of our country, we were not able to capture more data (of different genres) to further test and evaluate our method; the Greek folk and modern dances were under-trained with only 10 dances each.}. 
%

\AAAA{We have also experimented our approach with other dances genres, taken from the recently published AIST++ dance motion database~\cite{Li:2021}. Since the motions in this dataset have different skeletal proportions (.pkl format) than our data (.bvh format), our network has been re-trained explicitly with the AIST++ dance motions (in this experiment, we use $K = 150$ clusters). Note that this database is limited to short sequences that are less than 15 seconds (sBM) / 30 seconds (sFM) - comparing to performances that approximately last for 3 minutes - thus extraction of a meaningful global structure among similar dances is challenging. For that reason, the motif order was explicitly provided. As demonstrated in the accompanied video, our method was able to generate high-quality long dance sequences that follow both the rhythm, and the given choreographic rule.}
\subsection{Dataset and Implementation Details}
\label{subsection:Implementation details}
We invited 32 professional dancers and asked them to dance to different music tracks and recorded their dance using an optical motion capture system.
Our dataset consist of paired and synchronized dance to music performances, each lasts 60-180 seconds. We collected data for 3 different dance genres: approximately 2 hours of salsa dancing in couples (leader and follower), 20 minutes of modern (solo), and 20 minutes of folk dancing (solo), all with their corresponding music in different intensity and tempos (e.g., 140, 160, 180, 200, 220 bpm). The motion capture data were originally sampled between 120 to 480 frames per second (fps), but since human motion is locally linear, we reduce it to 30 fps without much loss of temporal information. The corresponding audio data are stereo, sampled at 44.1 KHz, with RMS normalization to 0dB using FFMPEG coding, at 30 fps. By default, the Librosa tool~\cite{Ellis:2007,McFee:2015} loads audio as mono (floating point time series), and downsamples audio rate to 22.05 KHz.

All of our experiments were run on an 8-core PC with Intel i7-6850K at 3.6GHz, 32GB RAM, and with NVIDIA Titan XP GPU. Training time is approximately 8 hours, and the testing (synthesis) is real-time. We have implemented our system in PyTorch using three fully connected layers with a memory size of 1024.

\begin{figure}[t]
	\includegraphics[width=\linewidth]{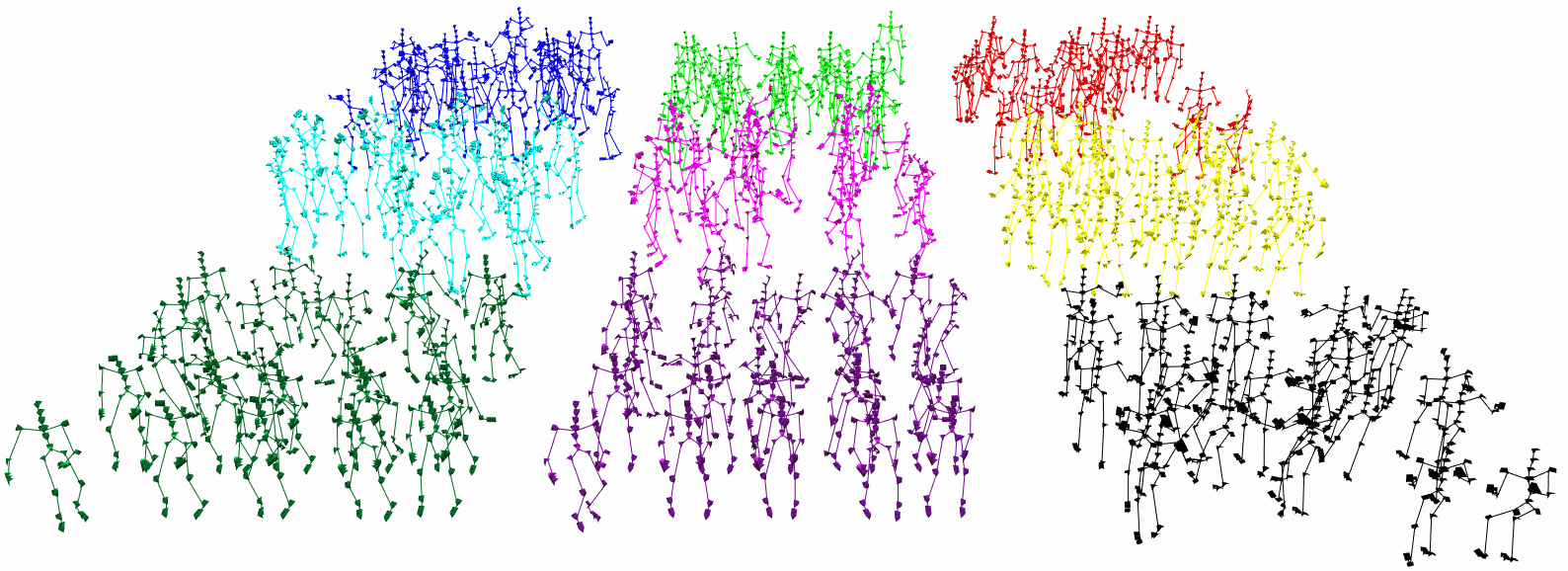}
	\caption{Motion words clusters. Our dataset consists of motion words belonging to clusters that are calculated in the motion embedding space. In this illustration, 9 clusters are visualized using 25 poses that were extracted from arbitrary motion words within each cluster.}
	\label{fig:Clusters}
\end{figure}

In our setup, we use the $J = 31$ most informative joints with their relative joint angles, represented in unit quaternions. Motion signatures is defined using $K = 500$ clusters, \AAAA{which includes motion motifs from all genres (salsa, Greek, and modern) together}. Experiments show that this number is sufficient to represent a diversity of movements within each cluster, while keeping the clusters compact with small mean error between motion words and their corresponding centroid. Figure~\ref{fig:Clusters} illustrates nine motion clusters, where each cluster is visualized by 25 poses that were extracted from arbitrary motion words within each cluster. \AAA{Recall that, during inference, our network is fed by 15 frames (poses), together with the motif that corresponds to these poses, and the foot contacts.} 

\subsection{Evaluation}
\label{subsection:Experimental Results}

Generating music-driven dance animations is a relatively new topic in character animation. This poses several challenges in the evaluation and comparison of our method, mainly due to the lack of available paired audio-to-dance motion datasets. Thus, we qualitatively evaluate our method, compared our methodology with baseline methods, and conducted ablation and observer survey for quantitative and qualitative evaluation. \AAA{The baseline method used in our experiments is the acLSTM network~\cite{Zhou:2018}, that was trained with the same dataset, but has been modified to take as input local joint rotations (represented by quaternions) instead of positions.} 
We used similar criteria to the 2D method of Lee~\etal~\cite{Lee:2019:dancing2music}, and added components that take into consideration the global content consistency of dance, and the control over the choreography. 

\subsubsection{Beat-to-Motion Synchronization}
\label{subsubsection:Beat-to-motion synchronization}
\begin{figure}[t]
	\includegraphics[width=\linewidth]{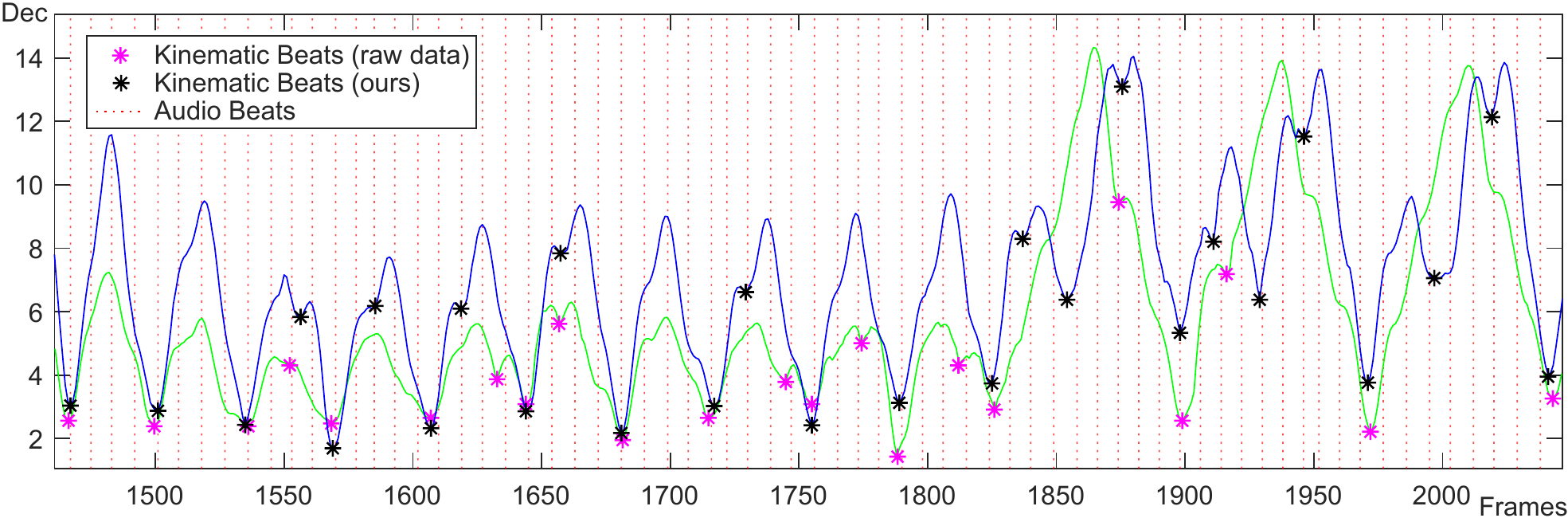}
	\caption{Music to dance synchronization. Kinematic beats (asterisks), which are represented by  local minima of joints deceleration (raw data in green line, ours in blue line) are synchronized with the acoustic beats (dotted red lines).}
	\label{fig:BeatSync}
\end{figure}

Dance is usually performed by periodic changes of the direction and/or speed of the motion that are synchronized with the musical beats. To demonstrate that our synthesized motion is synchronized to the beat, we compare the kinematic beat extracted from our motion with the corresponding input music beats. Kinematic beat is perceived as the sudden motion deceleration~\cite{Davis:2018}; in this work, we compute the Euclidean distance between two consecutive frames (the sum over all joint positions) as a function of time (raw data in green line, ours in blue), apply the Savitzky-Golay filter to remove noise, and then set kinematic offsets at the local minima. As visualized in Figure~\ref{fig:BeatSync}, most kinematic beats generated by our method (denoted by a black asterisk) are synchronized to the musical beats (dashed line), and are aligned to the kinematic beats of the raw motion capture data (shown in magenta asterisk) for the same music. A quantitative analysis on the beat-to-music synchronization is given later in subsection~\ref{subsection:Ablation Study}.

\subsubsection{Style Variation}
\label{subsubsection:Diversity}
\begin{figure}[t]
	\includegraphics[width=\linewidth]{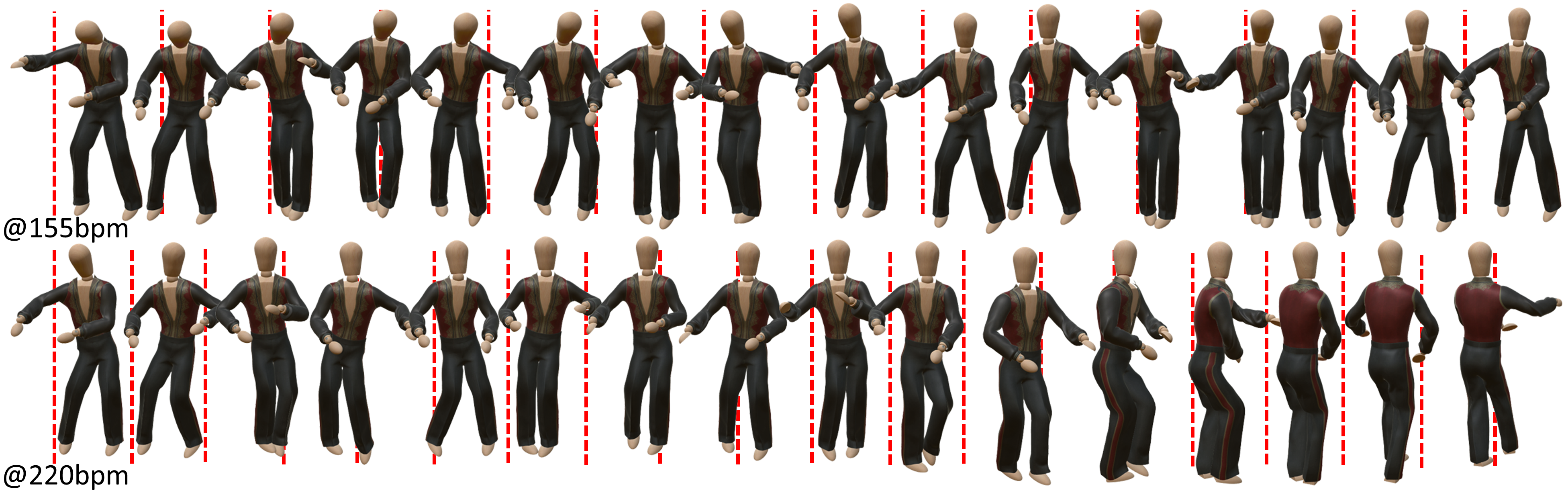}
	\caption{Beat per minute comparison. Our framework generates different motion sequences for music signals with different beat rate per minute (155bpm, and 220bpm), but still with the same target motifs sequence. Red dashed lines indicate the beats.}
	\label{fig:Style Diversion}
\end{figure}

To evaluate the style diversity of our generative model, we synthesize motion sequences, driven by the same target motifs, using music signals with different spectrum and beat per minute. Note that the rhythmic features itself add a rhythmic/tempo condition on the generation of motion that affect the style of the synthesized motion (e.g, pace, motion speed, the movement). Figure~\ref{fig:Style Diversion} shows selected frames from two motion sequences that were generated by our framework given music signals of 155 bpm (top) and 220 bpm (bottom), where the beat rate is highlighted in red dashed vertical lines. It can be seen that while following the beat, they perform the same choreography, following the target motifs, but in different phases and with a subtle variation of poses. The corresponding animations can be found in the supplementary video. 

\begin{figure}[t]
	\includegraphics[width=\linewidth]{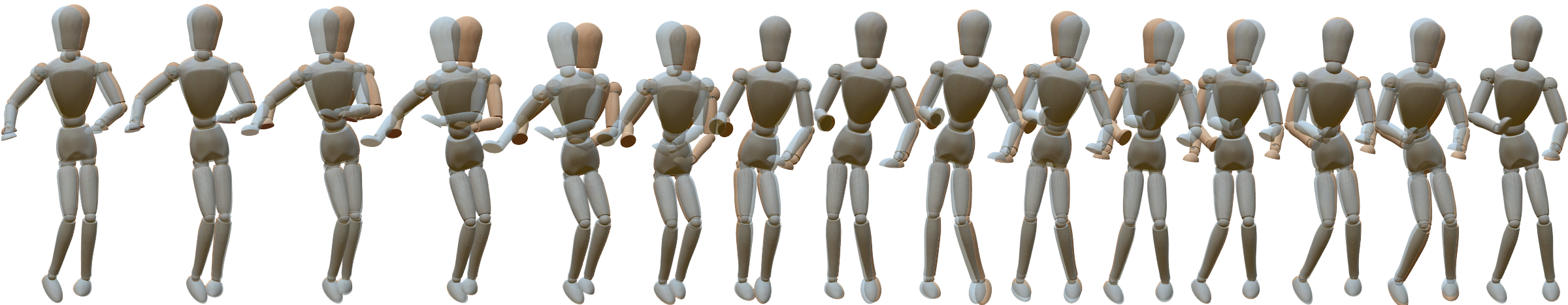}
	\includegraphics[width=\linewidth]{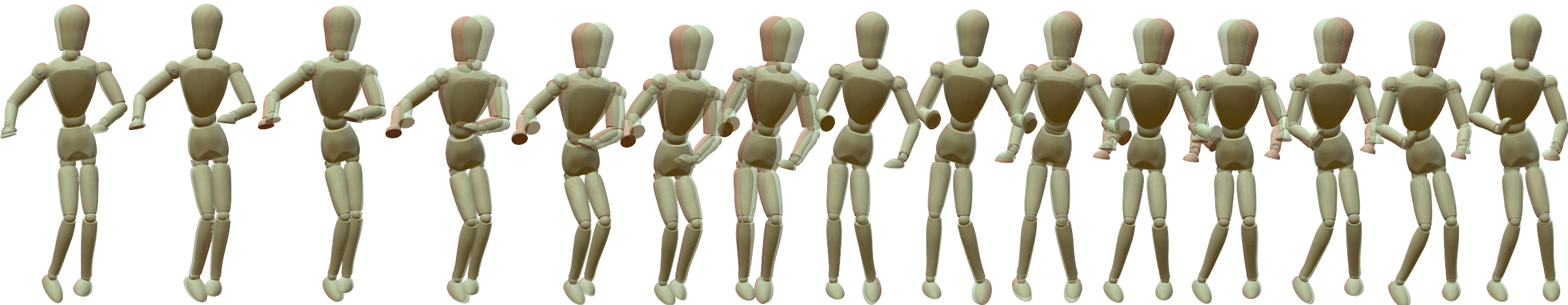}
	\caption{Style variation as a function of the spectral features. Three animations were generated with the same input parameters, except for the input spectral features. A sequence that was generated without spectral features (no AdaIN layer), overlaid by a sequence that was generated with spectral audio features, is shown in each row. The music tracks share the same beat-per-minute.	
	For an animated version, refer to the supplementary video.}
	\label{fig:Style Diversiity}
\end{figure}

Besides the rhythmic features, the spectral features also control the variation of the output motions through an AdaIN layer. To demonstrate this effect, we used a fixed input to the acLSTM network (rhythmic features, motifs) and performed 3 experiments. In the first, we did not inject style variation through the AdaIN layer, while in the other two we injected spectral features extracted from two different songs with the same bpm (166). Figure~\ref{fig:Style Diversiity} and the supplementary video show how the different output motions have subtle variations while their movements and kinematic beats are preserved. \AAAA{To further evaluate our style variation component, we conducted an ablation experiment, where we inject random noise to the system instead of the spectral audio features $\mathbf{a}_s^k$. This experiment examine whether this component can be controlled by signals that do not have the characteristics of spectral audio features. As can be seen in the supplementary video, the generated motion becomes noisy and unstable, which demonstrates that the AdaIN layer extract meaningful features that are pertinent to the spectral audio features.}

\subsubsection{Motion Variation}
\label{subsubsection: Motion Variation}
\begin{figure}[t]
	\includegraphics[width=\linewidth]{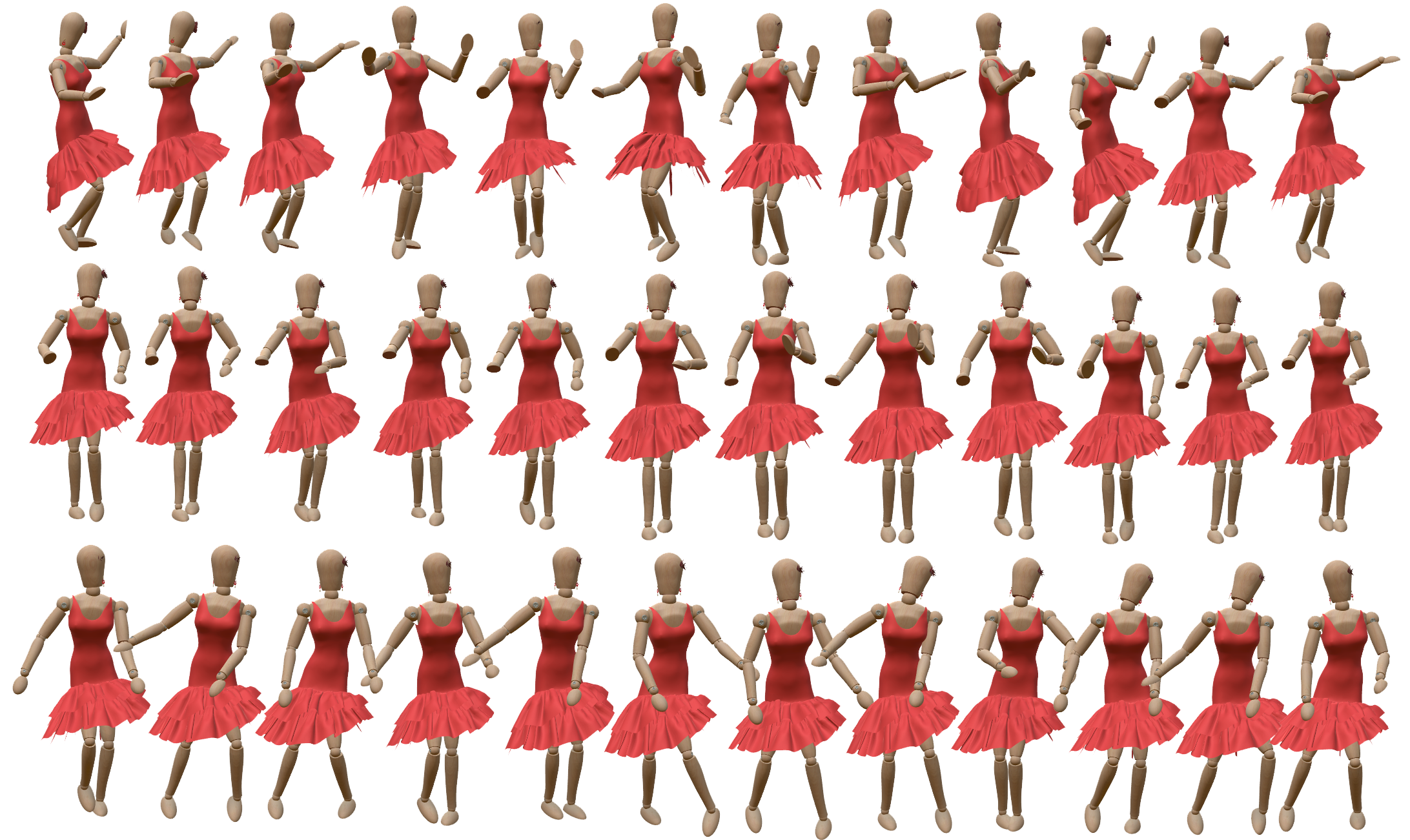}
	\caption{Motion variation. For the same target signature and music signal, our method produces three different choreographed dances. Each row shows 12 frames (between 440-550) that were extracted from the output sequences. Although different, the motions converge to the same target signature, respecting the global structure of the dance genre.}
	\label{fig:Multi Modality}
\end{figure}

To demonstrate the multi-modality and richness of our generated motions, we run our network several times with the same input music. \AAAA{In this experiment we used music samples that were taken from the training data}. Figure~\ref{fig:Multi Modality} shows selected poses from three sequences of dance (salsa follower) that were generated with the same inputs (music and target signature). Our method generates a different animation each time, but always converges to the template signature, proving that the global content for that dance genre remains coherent. In comparison, the standard acLSTM  always synthesizes the same dance. This is due to the controlled randomness in the selection of the target motifs by the choreography level. At each execution, a different dance is produced, exhibiting the rich variety and diversity in the content of the synthesized dance.

\subsubsection{Global Content Consistency}
\label{subsubsection:Global Content Consistency}
We next evaluate the global consistency of the content of our output with respect to the training dataset. To the best of our knowledge, this is the first method that enables preservation of global structure of the generated motions. The evaluation is done by quantifying the difference between the template motion signature (the average motion signature within the training dataset) to the signature of our generated data and to the data generated by the baseline.

To measure the distance between two signatures, we use the Chi-Square distance over the normalized signatures, given by
\begin{equation}
     d_s(\mathbf{s}_i, \mathbf{s}_j) = \frac{1}{2}\sum_{k=1}^{K} \frac{(\mathbf{s}_i^k - \mathbf{s}_j^k)^2}{(\mathbf{s}_i^k + \mathbf{s}_j^k)},
    \label{eq:chi_distance}
\end{equation}
where $\mathbf{s}_x^k$ is the $k$th component of signature $\mathbf{s}_x$. Table~\ref{tab:signature} reports the distances between signatures that were generated by our method and the baseline (acLSTM) to their corresponding template signature, for various dance genres. The signatures of the generated data using 10 examples for salsa leader and salsa follower, and with 3 examples for modern and Greek folk.

\begin{figure}[t]
\begin{tabular}{cc}
	\includegraphics[width=0.48\linewidth]{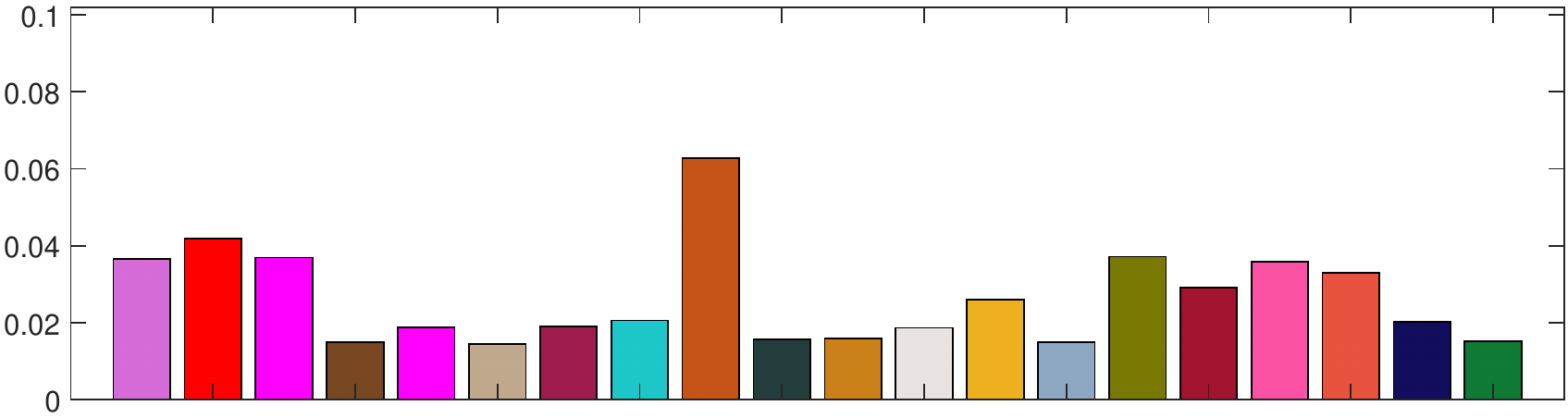} &
	\includegraphics[width=0.48\linewidth]{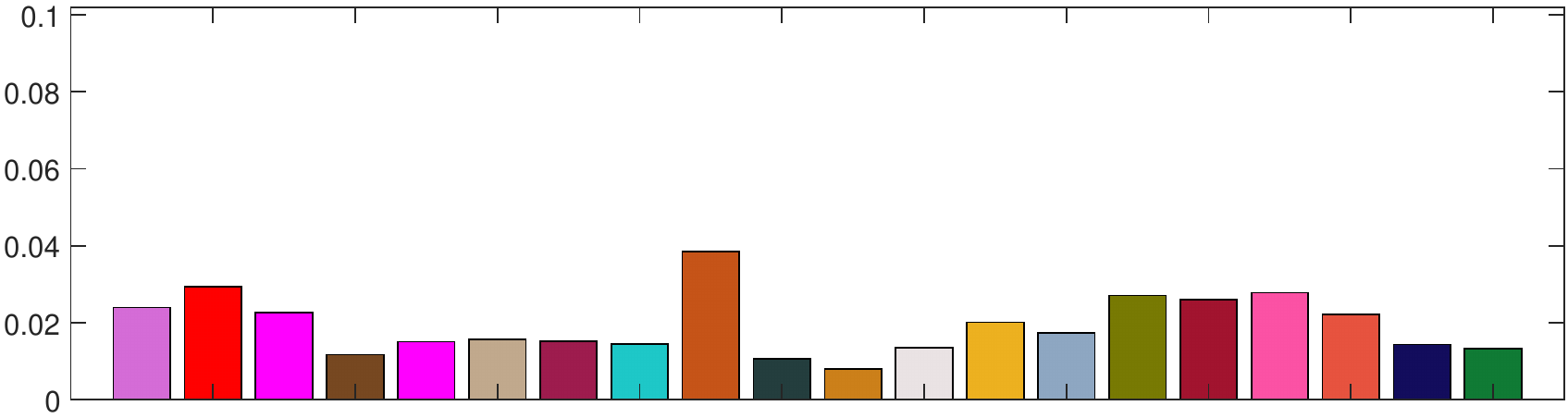} \\
	Template (and target) & Ours 1 \\
	\includegraphics[width=0.48\linewidth]{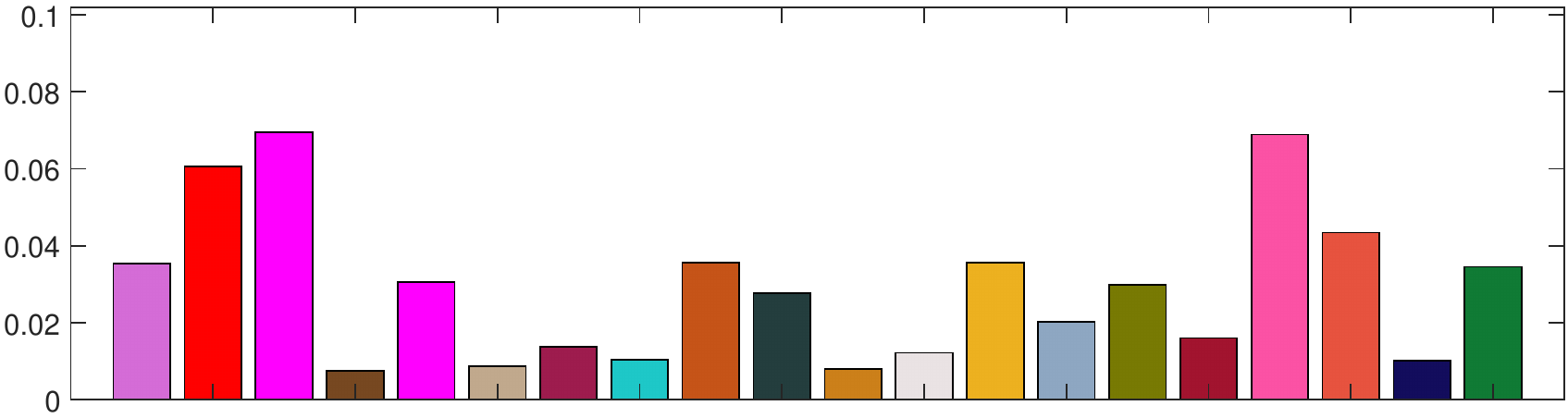} &
	\includegraphics[width=0.48\linewidth]{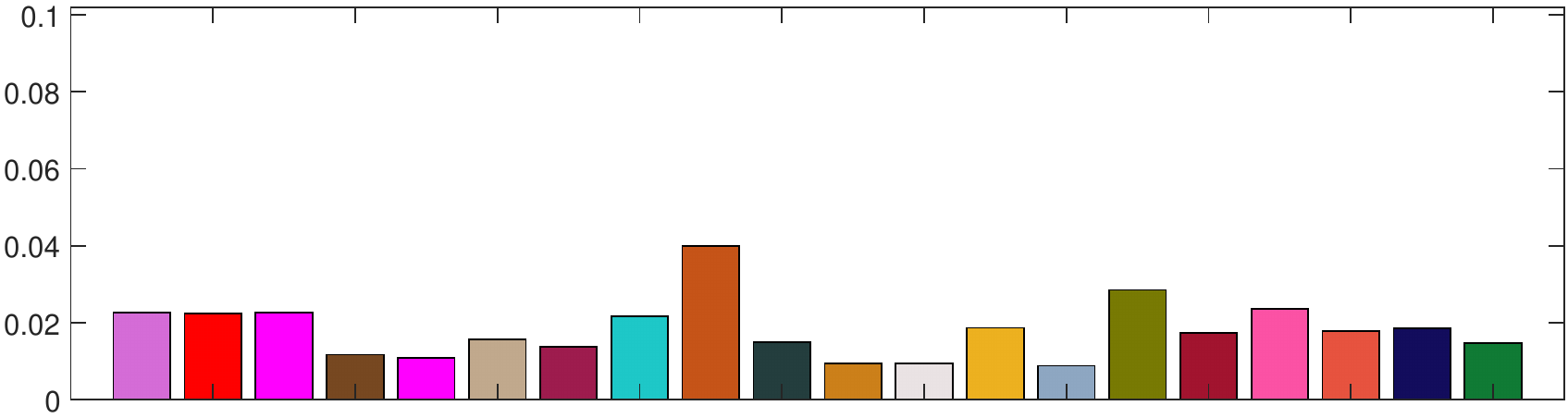} \\
	Raw data & Ours 2 \\
	\includegraphics[width=0.48\linewidth]{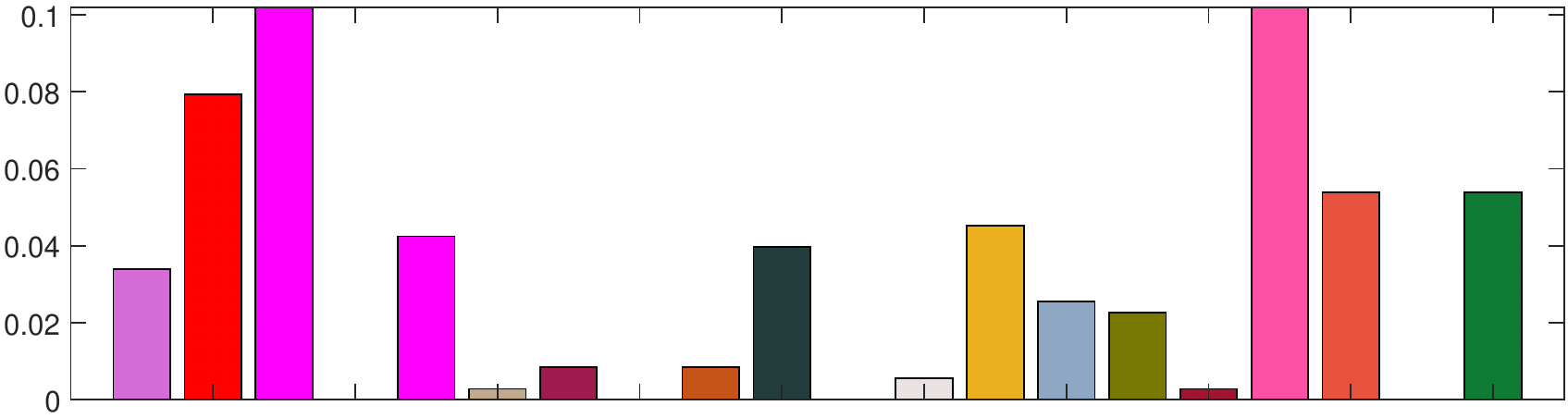} &
	\includegraphics[width=0.48\linewidth]{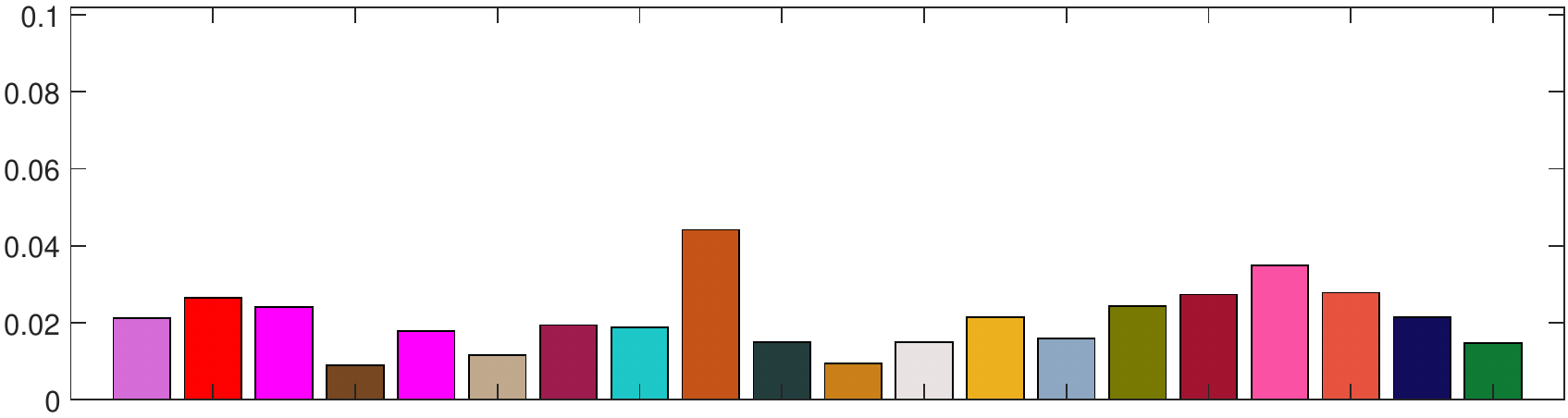}\\
	 acLSTM & Ours 3 \\
\end{tabular}
	\caption{Following a target motion signature (top left). On the left we show the raw data and baseline's (acLSTM) output signatures. On the right, 3 instances of signatures of our generated sequences using the specified target signature.}
	\label{fig:Motion Signatures}
\end{figure}

For a qualitative comparison, Figure~\ref{fig:Motion Signatures} visualizes 6 signatures (20 most frequent classes): a template signature, a signature generated with the baseline method, 3 signatures of different sequences that were generated by our method, and a signature of an arbitrary raw data dance (for reference). To examine reliable and rich global information, the motion signatures of our generated sequenced are extracted from clips of 250 motion words ($\sim 80$ [sec]). It can be seen that the signature of the acLSTM output demonstrates a mode collapse because of frequent repetitions of specific movements, while other movements tend to vanish. In contrast, the 3 arbitrary motion signatures of our generated sequences are converging to the template one.

\begin{table}[t]
    \centering
    \begin{tabular}{lccc}
    \toprule
                     & acLSTM & Ours (no $\mathcal{L}_{perc}$) & Ours \\
    \midrule
    Salsa (leader)   & 0.378 & 0.303 & 0.268 \\
    Salsa (follower) & 0.432 & 0.341 & 0.272 \\
    Greek Folk       & 0.345 & 0.296 & 0.240  \\
    Modern           & 0.371 & 0.337 & 0.323  \\
    \bottomrule
    \end{tabular}
    \caption{Quantitative evaluation of following a target motion signatures. Each value in the table represents the average distance between the motion signatures of the dance that was generated by the method specified in the columns to its target signature. Each row specifies a different dance genre. As can be seen, our method produces dances that are closer to the target signature.}
    \label{tab:signature}
\end{table}

\begin{figure}[t]
	\includegraphics[width=\linewidth]{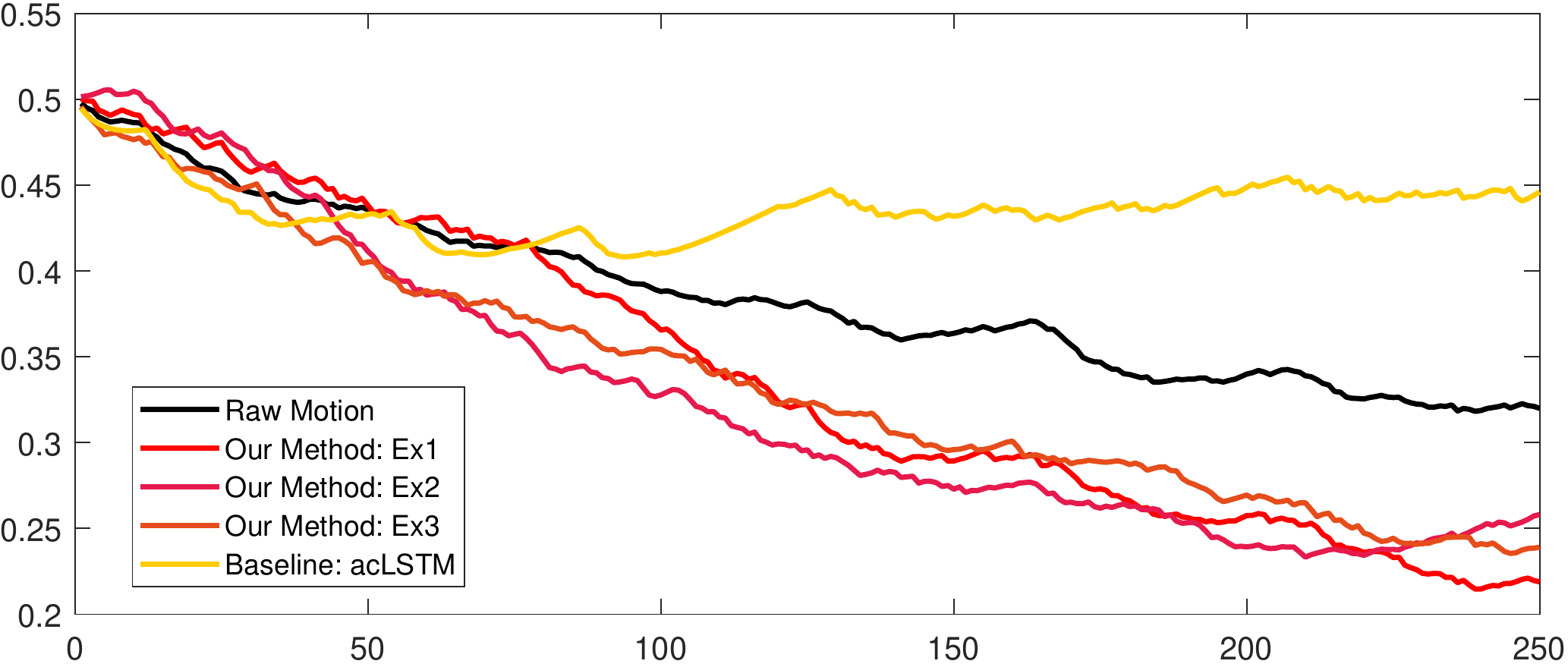}
	\caption{Distance of various signature to the template signature over time (the horizontal axis). The distance is calculated between the target signature and each of the following signatures in at constant time intervals: Raw motion (black), acLSTM (yellow), ours (red shaded colors).}
	\label{fig:Convergance_Graph}
\end{figure}

To visualize the convergence over time of the different signatures (in Figure~\ref{fig:Motion Signatures}) to the template signature (top-left of Figure~\ref{fig:Motion Signatures}), we computed the metric between these signatures and the template signature during their generation, at constant time intervals using the metric specified in \eqref{eq:chi_distance}.
Figure~\ref{fig:Convergance_Graph} shows the value of the metric over time. It can be observed that all of the samples generated by our method (red shaded colors) converge faster and are closer to the target signature than the raw data sample (black color). This because the target signature is the average signature over all the data sample, while specific dances are based on the improvisation of a specific performer. In addition, the baseline method (yellow color) can neither generate motions that respect the global structure of the dance genre nor ensure global consistency and variety in the generated data. Therefore, it does not converge to the template signature. 


\AAA{
\subsubsection{Choreography Mixer Evaluation}
\label{subsubsection: Mixer Evaluation}
To further evaluate our choreography mixer layer, we synthesize motion in the form of  
motion graphs~\cite{Kovar:2002}. In particular, the choreography mixer selects candidate motifs from the database; the system then transits to the motion word, from the corresponding cluster, that its first three poses have the smallest distance from the last three poses of the previous motion word. The selected motion word is time-scaled to match the given beat, and the two consecutive motion words are then stitched together. Note that, in this implementation (both for ours and motion graphs) the AdaIN network, and the IK post-processing have been disabled, while the motif order is explicitly provided so as both methods will perform the same dance. As it can be observed in the supplementary video, the generated motion using motion graphs looks temporally inconsistent, while motion word's blending adds foot sliding artifacts. When movements are common and not complex (thus clusters are compact), motion graphs perform better rather than when movements are complex and with rare motifs (clusters are loose). In addition, motion graphs do not have as much diversity as our network, and require large memory, which can be as large as tens of gigabytes, while our network compress data to the order of magnitude of megabytes (the weights of the neural network).}


\subsection{Observation Survey}
\label{subsection:observer survey}
We conducted an observational survey to evaluate some perceptual aspects of our results: beat-to-motion synchronization, realism, richness, and preservation of global structure of the dance genre. 45 observers participated in the survey, where 7 were professional dancers. The observers were asked to watch a series of short video clips that were coupled with questions related to our task. \AAAA{Note that we only used the salsa data for this observation study.} Readers are encouraged to watch the video in the supplementary materials which demonstrates a recorded user study answered by one of the subjects.

\begin{figure}[t]
	\includegraphics[width=\linewidth]{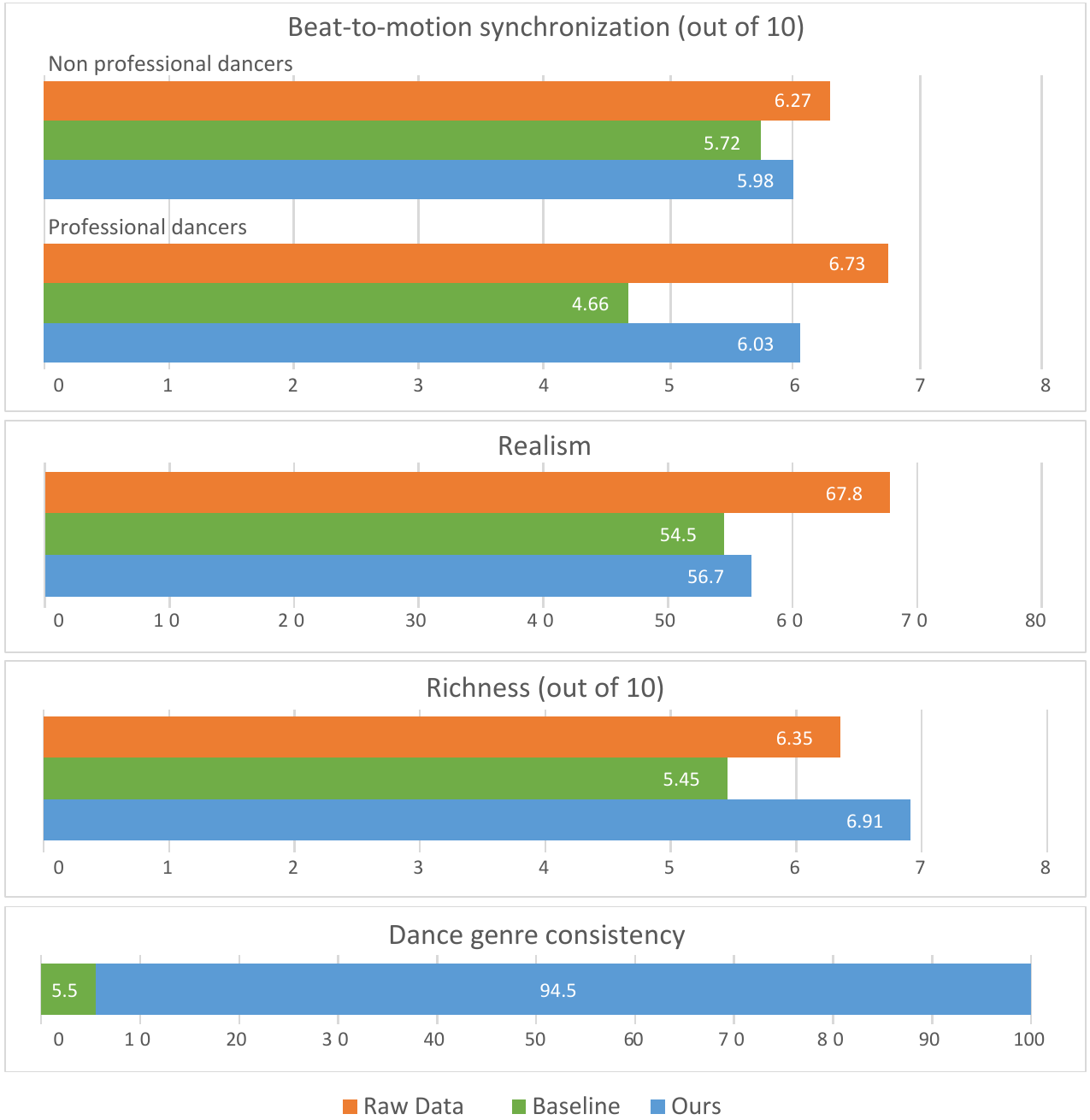}
	\caption{Our observation study, which exhibits the evaluation of four perceptual aspects of our generated results.}
	\label{fig:Observation_Study}
\end{figure}

\subsubsection{Beat-to-Motion Synchronization}
In this test, we asked the group of observers to evaluate ``How well is the dance motion synchronized with the music?'', by indicating a score between 0 to 10, where 0 means that the two are not synchronized at all, and 10 stands for perfect synchronization. Each observer watched 3 arbitrary clips, where one clip was extracted from the raw data motion clips, one from the baseline outputs and one from our results. Figure~\ref{fig:Observation_Study} lists the average score for raw motion, the baseline, and our method for two different groups of observers: non-professional and professional dancers. 
We conducted a paired t-test on the synchronization scoring between the baseline and ours, to show the importance of those results, and it returned $t = 2.34$ and $p = 0.0238$, indicating that the difference between the two is statistically significant. It can be seen that the group of non-professional observers struggled to distinguish between synchronized motions to non-synchronized ones, as the variance within the different ranks is small. In contrast, the group of professionals ranked the synchronization of our method distinctively higher than the baseline, which shows a clear improvement in the results.

\subsubsection{Realism}
\label{subsubsection:Realism}
The observers were next asked to perceptually evaluate the naturalness and realism of the generated dance motion with respect to the original motion capture data.
Note that for this set of clip questions we did not play the driving music with the motion, so the observers would only judge the realism of motion without being distracted by synchronization. The observers were presented with 6 raw and synthesized rendered motions, in random order. The participants were asked to mark the motion as one of two choices: ``A real motion captured data'', or ``A computer generated data''. Figure~\ref{fig:Observation_Study} exhibits the score for this experiment, indicating that the acLSTM baseline and our method received similar score in ``real'' labelling from the observers. It can be seen that the realism and naturalness of our generated dance motion cannot be easily distinguished from raw motion capture data. Moreover, the similar score that our method and the baseline one received for realism can be justified by the fact that the sequence of temporal coherent poses is generated (locally) with a similar technique that is used as the baseline. Note that for fare comparison, the baseline was trained with foot contact labels and went through post processing as well. 

\subsubsection{Richness}
For the same group of 6 motion clips that were presented in the realism test, the observers were asked to evaluate the richness of each motion clip, and to give it a score between 0 to 10, (0 -- ``not rich, mostly repetitive'', 10 -- ``very rich''). The average rank that the observer gave to the raw data, baseline and ours is shown in Figure~\ref{fig:Observation_Study}. 
We computed a paired t-test on the richness scoring between the baseline method and ours; the t-test returned $t = 2.96$ and $p = 0.0047$, indicating that the difference is statistically significant. It can be seen that our motions, which are intentionally driven to cover a wide range of movement within the dance genre received the highest score, even higher than the original raw data. 
In contrast, the baseline method that generates the sequence based (only) on a local prospective, outputs less richer and more monotonic motion sequences.

 \subsubsection{Consistency of Dance Genre}
To qualitatively evaluate how the global structure of the dance culture is preserved by our method, compared to the baseline, we presented a high level question that was shown only to the professional dancers, who can better evaluate high-level aspects of dance. The subjects watched 2 arbitrary pairs of clips, where within each pair, one is taken from our generated results, and the other displays the baseline results. The question was ``Which of the clips represent the Salsa genre better?''. As indicated in Figure~\ref{fig:Observation_Study}, 94.5\% of the results were in favor of our method, demonstrating the contribution of our motif, signatures, and choreography level.

\subsection{Ablation Study}
\label{subsection:Ablation Study}
To examine the performance and contribution of each hierarchy level in our implementation, we trained several versions of our network where in some we have removed the specific components that are examined.

\subsubsection{Audio Features}
To quantitatively evaluate the contribution of the injected audio features, we trained a model that discards $\textbf{a}_r$ and compared the synchronization of its outputs to our output and to raw data. We used two measurements; (a) the ratio of kinematic beats to musical beats, and (b) the ratio of aligned kinematic beats to the total kinematic beats. A kinematic and a musical beat are considered aligned if their corresponding distance is not larger than one frame. The ratios for our network without feeding the audio features are 24.1\% and 46.6\%, while our full framework (with the audio features) yields 24.4\% and 64.7\%. As a reference, we calculated the average of the ratios of our raw motion capture data samples (which are paired with music signals) and received 25.2\% and 72.5\%. These numbers indicate that the frequency of occurrence of a kinematic beat relative to that of a musical beat for our method is similar to those of raw data, meaning that the tempo of the generated dance is also similar, while the inclusion of the audio features in the training of the network shows significant improvement in the synchronization of the kinematic and musical beats. It is, thus, clear that our method exploits the information embedded in the audio features to align the synthesized motion with the input music beat, in almost the same manner as the synchronization of the raw motion capture data. 
 


\subsubsection{Foot Contact Labels}
Our method, in general, achieves temporally coherent and visually plausible results at the pose level. However, the foot contact awareness, and post processing, further improve the quality of the generated motions, eliminating noticeable visual foot sliding artifacts. The contribution of this step is demonstrated in the supplementary video.
\begin{figure}[t]
	\includegraphics[width=0.8\linewidth]{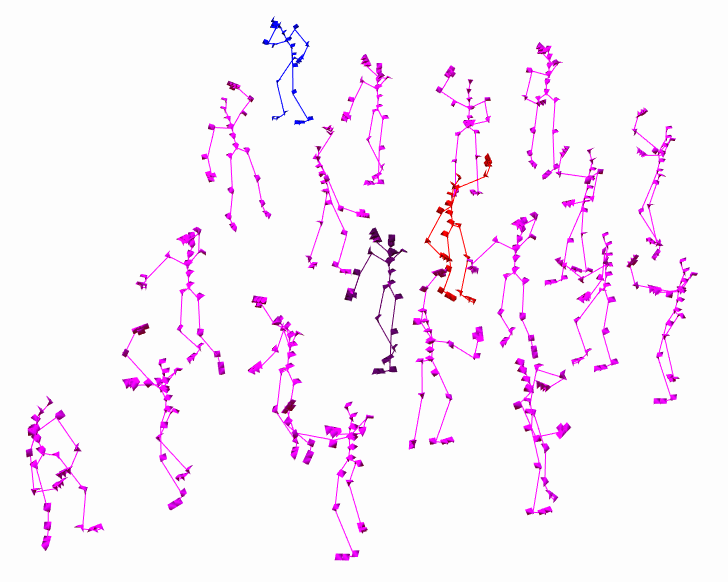}
	\caption{The importance of motion perceptual loss. The generated motion word in a system that was trained with the perceptual loss (red) is closer to the motif (dark purple at the center of its cluster of motions) compared to a motion word generated by a system trained without it (blue). The 2D embedding of motion vectors in this example created using the t-SNE algorithm.}
	\label{fig:EmbeddingSpacePrecLoss}
\end{figure}

\subsubsection{Motion Perceptual Loss}
In order to demonstrate the effect of the perceptual loss, we trained our framework on different datasets (different dance genres) without this loss. Table~\ref{tab:signature} reports the distances between the target signatures to signatures that were generated by the model without the perceptual loss (middle column) and the model with perceptual loss (right column), for the various genres. It can be seen that the presence of the motion perceptual loss, which encourages the generated motion words to be closer to their corresponding target motif in the motion embedding space, affects the generated motion signatures. The resulting signatures without perceptual loss are less similar to the template signatures compared to the ones generated by our full trained framework. To visualize this effect, Figure~\ref{fig:EmbeddingSpacePrecLoss} displays an example of a motion word that is closer to its motif if generated with motion perceptual loss.

\subsection{Applications}
\label{subsection:Application}

One of the main features of our method is that it allows control over the generated motion, using the choreography level, while ensuring the global consistency of dance. More specifically, the use of perceptual loss allows authoring control over the content of the generated motion. Such a control facilitates the adaptation and customization of the dance. It can be used to assist choreographers to complete their creative vision, granting them with the tools to compose specific dance scripts, or complete half-finished scenarios.

In this section, we present two applications that demonstrate the effectiveness of our method in controlling the global content of the generated dance. The main difference between our original method and these applications is that the choreography control is guided by another motion sequence or provided by the user.  

\subsubsection{Re-creation of a Choreography}
\label{subsubsection:Recreate the dance}
This application highlights how our method allows control over the specific global structure of the generated motion. Instead of using our choreography-level to generate the sequence of motion motifs at random (according to the signature and the motif transition matrix), we can supply the sequence of motifs from an outside source (like a real choreographer). 

\begin{figure}[t]
	\includegraphics[width=\linewidth]{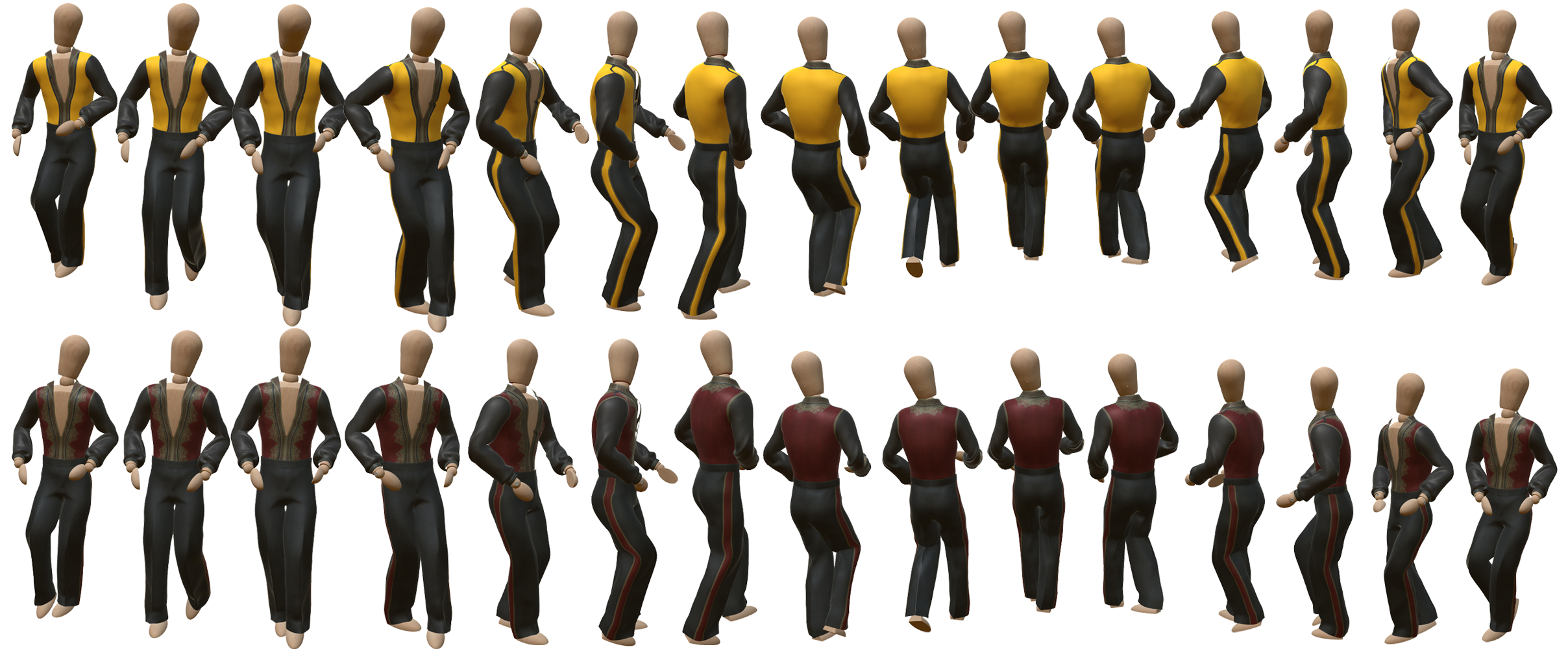}
	\caption{ Motion regeneration. Our framework can imitate an existing dance performed by a professional dancer. The sequence of motion motifs are extracted from the original dance (top) and passed to our network to guide the synthesis and generate a similar dance (bottom). See also the supplemental video.}
	\label{fig:Application1}
\end{figure}

To demonstrate this ability, we extract a sequence of motion motifs from an existing dance of a professional dancer that is not part of the training set. We use this motif sequence as the driving sequence of our method by disabling the choreography layer, and synthesize a dance to the same music.
Figure~\ref{fig:Application1} shows the original dance performed by a professional dancer (top), and the reconstructed dance generated using our method (bottom). Please refer to the accompanied video for the animated version. It is clear that our framework leads the network to generate motion that follows the movement of the original dancer. However, since a motif defines a cluster of movements, rather than a specific movement, it can be seen that the generated sequence does not identically imitate the original one. For example, in some parts, the limbs may have different orientations, or the global orientation of the skeleton may be different due to the fact that orientation is not part of the evaluation of the motif which is described by local joints. In addition, it can be observed that the parts that are mostly correlated, are the ones that contain common movements (e.g., the salsa styles: mambo and cucaracha). 

%

\subsubsection{Filling the Gaps}
\label{subsubsection:Filling the gaps}
It may be difficult to compose a whole dance sequence using motifs. However, to explore possible variations, a choreographer may want to indicate specific motion motifs at specific times of the dance. 
We have implemented a tool that enables a user to indicate the desired motifs at specific times, and our method fills the missing gaps between these words, with a motion that is globally consistent and can follow a given genre signature.

\begin{figure}[t]
	\includegraphics[width=\linewidth]{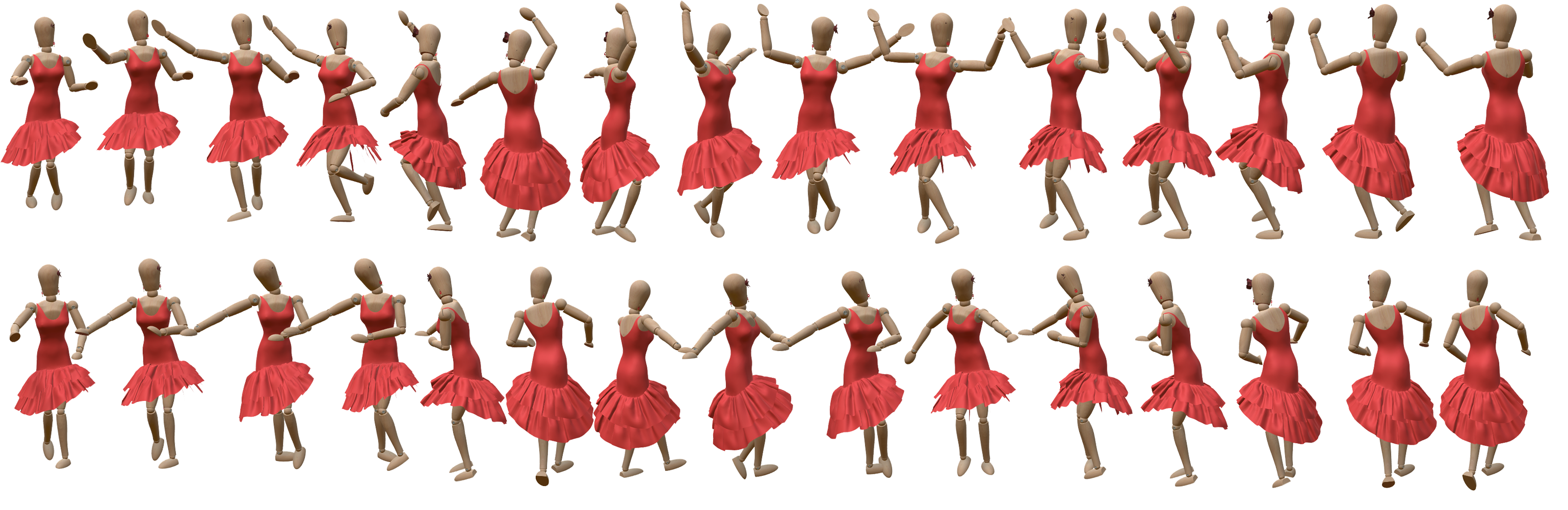}
	\caption{Diversity while filling the gap. Our method generates diversity within a motion cluster and does not simply follow the existing motion words. We show the input motif that represent a right turn with arms up (top), and our generated movement -- a right turn with arms down (bottom), which is not identical.}
	\label{fig:Application2}
\end{figure}

More specifically, the user selects specific dance motifs, at certain times; these motifs are set as constraints that drive and control the animation at these specific times. Whenever they are not specified, our 3D choreography model generates dance movements by generating the motif sequence in the usual manner, following the target signature and motif transitions.
Figure~\ref{fig:Application2} shows an example of motion synthesis that follows the constrained motif that has been provided by the user. In between the constrained time intervals, our method fills the gaps. This example can also be seen in the supplemental video, and clearly demonstrates a direct method to control the diversity of movements within the generated motion sequence. 

Note that this operation does not correspond to classic motion in-betweening, that enables generation of smooth transitions between two key-frames. Unlike the latter, here we generate a sequence of poses that starts and ends with motion words correspond to the specified start and end motifs, satisfying the global contextual structure.

%


\section{Discussion, Conclusions \& Future Work}
\label{section:Conclusions and Future Work}

In our work, we have addressed the challenging problem of generating long-term, temporally coherent movements with style diversity which, at the same time, preserve the global structure of a given choreography. 
The key feature of our framework is that it enables controlling and composing short-term motion words representing certain motifs into long-term sequences that adheres to target global distribution. As a result, we have achieved a generative model of dances with meaningful and rational structures. 
Except for generating a coherent sequence of motion words with a global distribution of motifs, the sequence also follows an audio stream synchronously, using the audio features to drive the animation. The generated dance is aligned with the music rhythm, but also presents richness in the style of movements.
In our results, we have demonstrated the importance of having control over the global structure of the dance in several applications, and evaluated the performance of our network in terms of realism,  synchronization to the beat, style diversity, and multi-modality. It is important to note that this framework is not limited on subjects related to digital choreography, but it can find also applications in other domains, where the content is characterized by a combination of sub-actions, e.g., sports, martial arts.

\subsection{Limitations} Our method has limitations. First, our training motion captured data has some artifacts, mainly due to errors during data acquisition, that also affect the quality of the synthesized motion. \AAA{Other acoustic features, such as the volume, melody, or different instruments could also be taken into consideration.} The number of clusters that define the motifs plays an important role in synthesis, more particularly it affects the convergence of the network, which is highly depended in the training dataset. \AAA{Even though we used a large dataset of dance motions, there are still motions not represented in our data, and therefore, in our universal feature space. This means that the number of clusters may need to be tweaked for different dance genres or datasets. Similarly to the original motion signatures work~\cite{Aristidou:2018}, another limitation incurs on the building of the signature where all motion words are clustered to the closest motifs. In that manner, outliers cannot be correctly assigned to their representative cluster, thus the accuracy of the signature created is reduced; in addition, it makes clustering prone to averaging effects, especially if the diversity of the motion words is large, or the outlier word is far away from the centroid.} We also encountered a number of outliers in the embedding space mainly due to erratic motions in the training data.



Although our system provides control over the composition of the choreography, it cannot replace the choreographer's creativity, which includes artistic concepts such as spontaneity, improvisation, experience, and talent. Further studies are required to create autonomous dancers that are capable of composing novel motifs or original dance choreography scenarios, based on their own improvisation.

\subsection{Future Directions} In the future, we aim to train a generative model on a larger dataset that includes a variety of dance types and motions, different music styles etc. \AAA{In addition, when forming the average motion signature for each dance genre, the standard deviations for the proportion of each motif could be used to create more realistic dance variations.}

In a wider perspective, our work introduces a rather unique generative model that combines global statistics into the synthesis of a locally coherent stream. We believe that similar principles can be applied to other media, for example synthesizing images or videos that follow some global statistics, like color distribution or target semantic meaning.
Another future avenue of research we would like to consider the combination of two (or more) virtual dancers interacting. The driving motif on each beat can be selected based on feedback gained from the interactive character. A challenging problem is to allow one virtual dancer to improvise and the other to react respectively to keep the generated dance coherent. This can also allow interaction between real and virtual performers.

\appendices
\section*{Appendix}
\label{ref:appendix}

\begin{table}[h]
    \centering
    \caption{The audio features used in our implementation. For more details, please refer to the Librosa audio features description~\cite{McFee:2015}.}
    \begin{tabular}{c|l}
    \multicolumn{2}{l}{\emph{Rhythmic features}}   \\
    \midrule
        $a_1$ &  Tempogram: local autocorrelation of the onset strength \\    &  envelope. \\
        $a_2$ & Spectral flux onset strength envelope \\
        $a_3$ & Root mean square value: to approximate downbeats.\\
        $a_4$ & Beat rate (binary form) \\
    \multicolumn{2}{l}{ \ } \\
    \multicolumn{2}{l}{\emph{Spectral features}} \\
    \midrule
        $a_{5-16}$ & Chromagram: 12-element feature vector \\ 
        $a_{17-28}$ & Constant-Q chromagram: 12-element feature vector. \\
        $a_{29-40}$ & Chroma Energy Normalized: 12-element feature vector\\
        $a_{41-66}$ & Mel-scaled spectrogram: 26-element feature vector\\
        $a_{67-79}$ & Mel-frequency cepstral coefficients: 13-element feature \\ & space \\ 
        $a_{80}$ & Spectral centroid \\
        $a_{81}$ & Spectral bandwidth \\
        $a_{82}$ & Spectral contrast \\
        $a_{83}$ & Spectral flatness \\
        $a_{84}$ & Spectral Roll-off frequency\\
        $a_{85-90}$ & Tonal centroid features\\
        $a_{91}$ & Zero-crossing rate\\
    \end{tabular}
    \label{tab:audio features}
\end{table}


\ifCLASSOPTIONcompsoc
  \section*{Acknowledgments}
\else
  \section*{Acknowledgment}
\fi
This work has received funding from the University of Cyprus. It has also received funding from the European Union's Horizon 2020 Research and Innovation Programme under Grant Agreement No 739578 and the Government of the Republic of Cyprus through the Deputy Ministry of Research, Innovation and Digital Policy.

\ifCLASSOPTIONcaptionsoff
  \newpage
\fi



\bibliographystyle{IEEEtran}
\bibliography{References}

%



%

\begin{IEEEbiography}[{\includegraphics[width=1in,height=1.25in,clip,keepaspectratio]{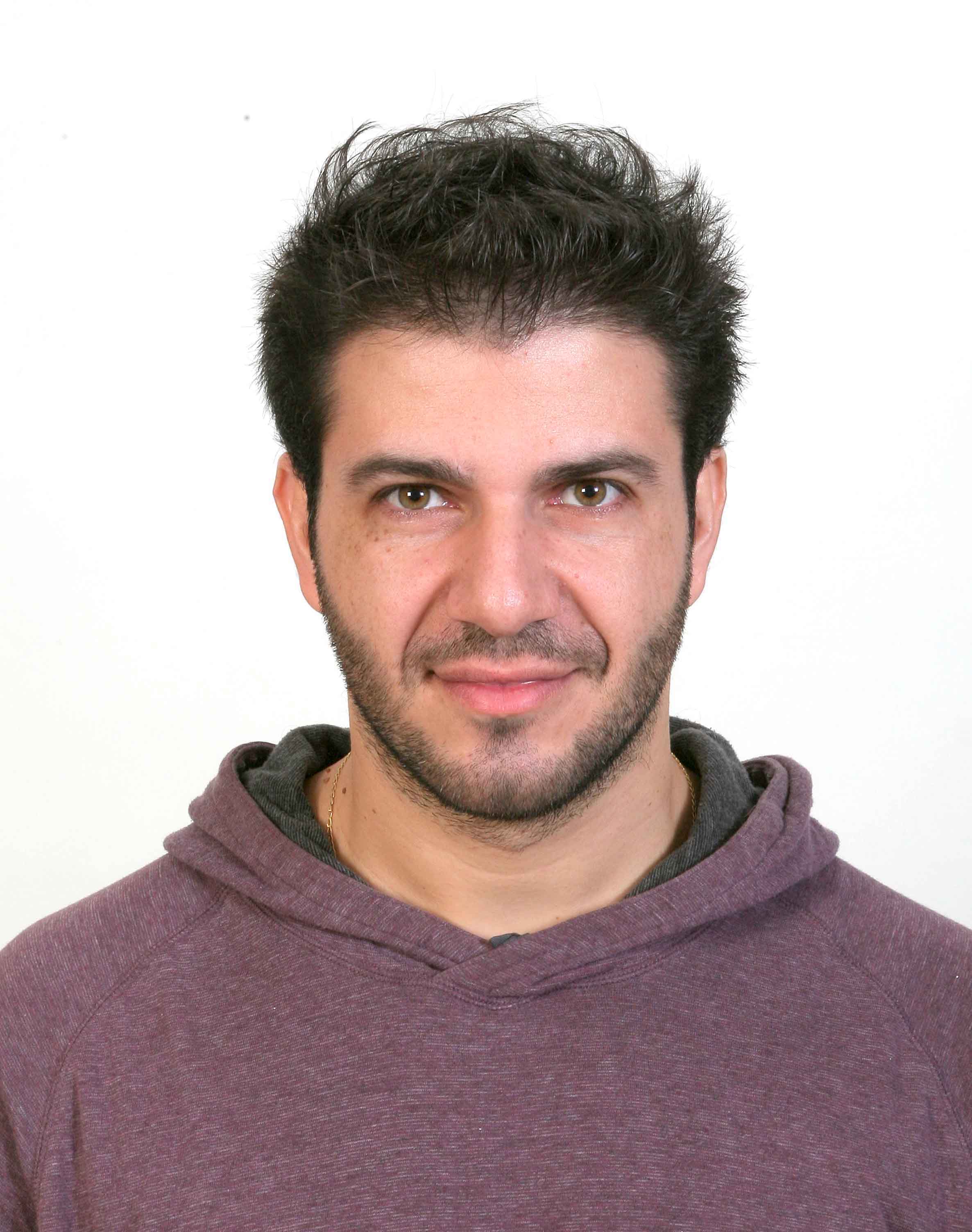}}]{Andreas Aristidou} is an Assistant Professor at the Department of Computer Science, University of Cyprus. He had been a Cambridge European Trust fellow at the University of Cambridge, where he obtained his PhD. Andreas has a BSc in Informatics and Telecommunications from the National and Kapodistrian University of Athens and he is an honor graduate of Kings College London. His main interests are focused on character animation, motion analysis, synthesis, and classification, and involve motion capture, inverse kinematics, deep and reinforcement learning, intangible cultural heritage, and applications of Conformal Geometric Algebra in graphics.
\end{IEEEbiography}


\begin{IEEEbiography}[{\includegraphics[width=1in,height=1.25in,clip,keepaspectratio]{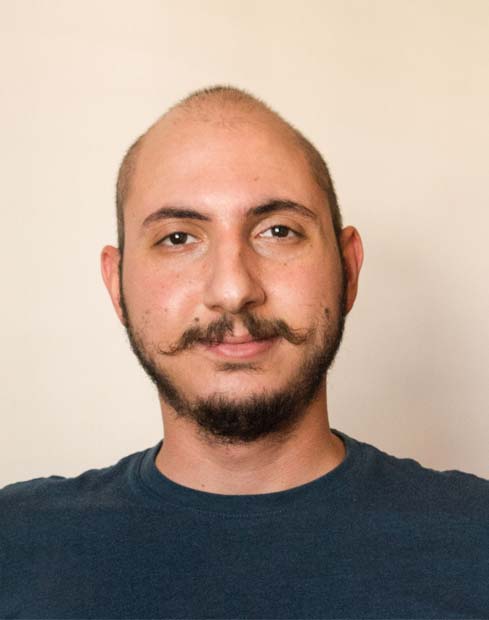}}]{Anastasios Yiannakidis}
is a Research Scientist at the Department of Computer Science, University of Cyprus. Anastasios received a BSc in Computer Science (honors) at the University of Cyprus, and since 2018, he works on projects related to 3D character animation. His interests are focus on motion synthesis and skeletal reconstruction, and involves deep and convolutional learning.
\end{IEEEbiography}

\begin{IEEEbiography}[{\includegraphics[width=1in,height=1.25in,clip,keepaspectratio]{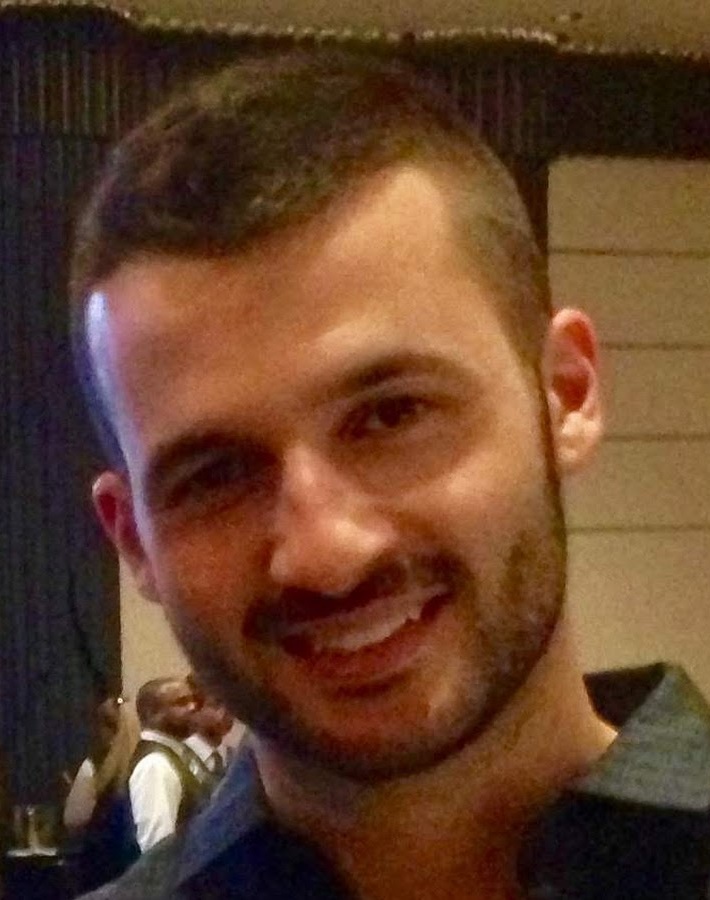}}]{Kfir Aberman}
is a research scientist at Google Research in San Francisco. His research interests include deep neural network architectures for various computer graphics applications. In particular, his work is focused on analysis, synthesis, and manipulation of human motion in real videos as well as 3D character animation. Kfir received his PhD from the Electrical Engineering department at Tel-Aviv University, and his MSc (Cum Laude) and BSc (Summa Cum Laude) from the Technion. He serves as a reviewer of various journals and conferences within the graphics community.
\end{IEEEbiography}

\begin{IEEEbiography}[{\includegraphics[width=1in,height=1.25in,clip,keepaspectratio]{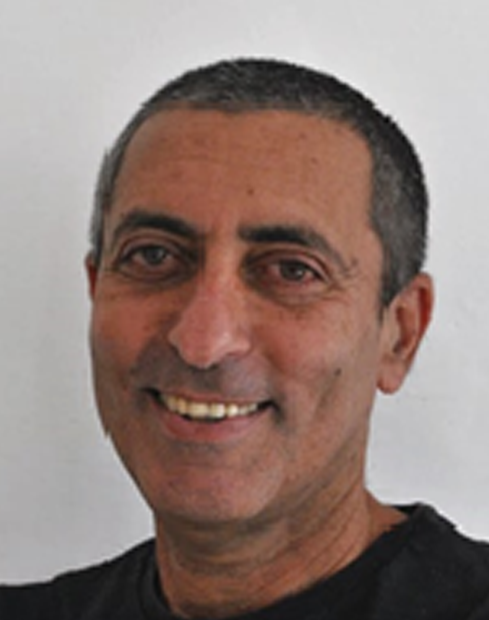}}]{Daniel Cohen-Or}
received the BSc degree in both mathematics and computer science, in 1985, the MSc degree in computer science from Ben-Gurion University, in 1986, and the PhD degree from the Department of Computer Science, State University of New York at Stony Brook, in 1991. He is a professor with the Department of Computer Science, Tel Aviv University. His research interests include computer graphics, visual computing and geometric modeling and including rendering and modeling techniques, shape analysis, shape creation and editing, 3D reconstruction, photo processing, compression and streaming techniques, visibility, point set representation, morphing, and volume graphics.
\end{IEEEbiography}

\begin{IEEEbiography}[{\includegraphics[width=1in,height=1.25in,clip,keepaspectratio]{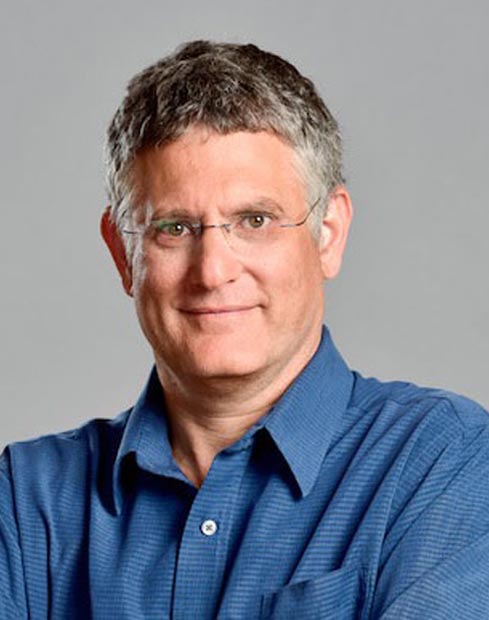}}]{Ariel Shamir}
is the Dean of the Efi Arazi school of Computer Science at the Interdisciplinary Center in Israel (now Reichman University). He received his PhD in computer science in 2000 from the Hebrew University in Jerusalem, and spent two years as PostDoc at the University of Texas in Austin. Prof. Shamir has numerous publications and a number of patents, and was named one of the most highly cited researchers on the Thomson Reuters list in 2015. He specializes in computer graphics, image processing and machine learning. He is a member of the ACM SIGGRAPH, IEEE Computer, AsiaGraphics and EuroGraphics associations.
\end{IEEEbiography}

\begin{IEEEbiography}[{\includegraphics[width=1in,height=1.25in,clip,keepaspectratio]{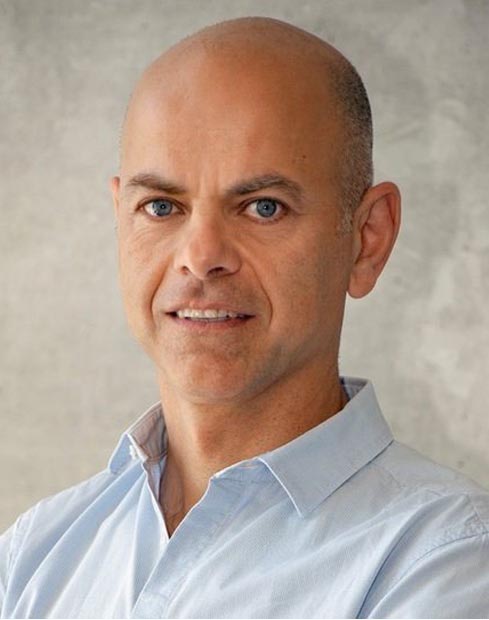}}]{Yiorgos Chrysanthou}
is a Professor at the Computer Science Department of the University of Cyprus where he is heading the Graphics and Hypermedia lab. He is also the Research Director of the CYENS - Centre of Excellence. Yiorgos was educated in the UK (Queen Mary College, University of London) and worked for several years as a research fellow and a lecturer at University College London. His research interests lie in the general area of 3D Computer Graphics, computer animation, algorithms for real-time AR and VR rendering and reconstruction of urban environments.
\end{IEEEbiography}




\end{document}